\documentclass{ldbc}

\usepackage{multirow}
\usepackage{float}
\usepackage{amsfonts}
\usepackage{bbding}
\usepackage{xcolor}
\usepackage{rotating}
\usepackage{pifont} 
\usepackage{amsmath}
\usepackage{lscape}
\usepackage{longtable}
\usepackage{listings}
\usepackage{hyperref}
\usepackage{fancyvrb}
\usepackage{subcaption}
\usepackage{algorithm}
\usepackage{algpseudocode}
\usepackage{pbox}
\usepackage{listings}
\usepackage{enumitem}
\usepackage[minnames=1,maxnames=3]{biblatex}
\usepackage[scale=0.78,ttdefault=true]{AnonymousPro}
\usepackage{todonotes}

\presetkeys%
    {todonotes}%
    {inline,backgroundcolor=yellow}{}

\LetLtxMacro{\oldTodo}{\todo}
\renewcommand{\todo}[1]{\oldTodo{TODO: #1}}

\bibliography{ms}

\definecolor{eclipseStrings}{RGB}{42,0.0,255}
\definecolor{eclipseKeywords}{RGB}{127,0,85}
\colorlet{numb}{magenta!60!black}

\lstdefinelanguage{json}{
  basicstyle=\footnotesize\ttfamily,
  commentstyle=\color{eclipseStrings}, 
  stringstyle=\color{eclipseKeywords}, 
  numbers=none,
  numberstyle=\scriptsize,
  stepnumber=1,
  numbersep=8pt,
  showstringspaces=false,
  breaklines=true,
  frame=lines,
  string=[s]{"}{"},
  comment=[l]{:\ "},
  morecomment=[l]{:"},
  literate=
    *{0}{{{\color{numb}0}}}{1}
      {1}{{{\color{numb}1}}}{1}
      {2}{{{\color{numb}2}}}{1}
      {3}{{{\color{numb}3}}}{1}
      {4}{{{\color{numb}4}}}{1}
      {5}{{{\color{numb}5}}}{1}
      {6}{{{\color{numb}6}}}{1}
      {7}{{{\color{numb}7}}}{1}
      {8}{{{\color{numb}8}}}{1}
      {9}{{{\color{numb}9}}}{1}
}

\newcommand{\toolname}{Graphalytics}%

\newif\ifrelease 

\releasetrue

\ifrelease
  
  \newcommand{\futureinversion}[2]{{}}
\else
  
  \newcommand{\futureinversion}[2]{{\color{red}#2 (Future in v.#1)}}
\fi

\newcommand{\alert}[1]{\textit{\textbf{{\color{red}#1}}}}

\WPTitle{Graphalytics team}

\renewcommand{\delIDText}{}
\delName{The LDBC Graphalytics Benchmark \\ v\lastVersionText{}}

\dueDate{M12}
\submissionDate{M13}

\newcommand{\mainmatterheadingl}{\relscale{.92} The LDBC Graphalytics Benchmark (v\lastVersionText{})}

\dissPU 

\natR 

\author{Alexandru Iosup (Vrije Universiteit Amsterdam (VU) and Delft University of Technology (TUD))}
\authorPartner{Alexandru Iosup (VU/TUD)}
\responsibleAuthor{Alexandru Iosup}
\responsiblePartner{VUA/TUD}
\responsiblePhone{}
\responsibleEmail{a.iosup@vu.nl}

\contributor{%
Ahmed Musaafir (VU),
Alexandru Uta (VU),
Arnau Prat P\'{e}rez (UPC),
G\'{a}bor Sz\'{a}rnyas (MTA-BME, CWI),
Hassan Chafi (Oracle),
Ilie Gabriel T\u{a}nase (IBM),
Lifeng Nai (GT),
Michael Anderson (Intel),
Mihai Capot\u{a} (Intel),
Narayanan Sundaram (Intel),
Peter Boncz (CWI),
Siegfried Depner (formerly Oracle),
Stijn Heldens (UTwente, formerly TUD),
Thomas Manhardt (Oracle),
Tim Hegeman (TUD),
Wing Lung Ngai (VU, formerly TUD),
Yinglong Xia (Huawei)
}

\reviewerOne{\alert{???}}
\reviewerTwo{\alert{???}}

\keywords{Graphalytics, graph analytics, benchmark}

\versionLog{
  \versionLogEntry{xx/yy/2019}{1.0.0}{}{Spelling corrections, improved consistency and the the presentation of the algorithms, added visualization of validation graphs}
  \versionLogEntry{28/06/2017}{0.9.0}{}{Providing the first stable release}
  \versionLogEntry{11/05/2017}{0.2.12.2}{}{Updating benchmark process and competition}
  \versionLogEntry{11/05/2017}{0.2.12.1}{}{Describing primary/secondary/other metrics, and participation in Graphalytics competitions}
  \versionLogEntry{07/09/2016}{0.2.7}{}{Describing result format and scoring mechanism}
  \versionLogEntry{07/09/2016}{0.2.6}{}{Minor fixes by Arnau for spelling/inconsistencies}
  \versionLogEntry{04/09/2016}{0.2.5}{}{First draft}
  \versionLogEntry{02/09/2016}{0.2.4}{}{First draft}
  \versionLogEntry{16/07/2016}{0.1}{}{First draft}
}
\lastVersion{1.0.5}

\documentUrl{}

\abstract{
In this document, we describe LDBC Graphalytics, an industrial-grade benchmark for graph analysis platforms. 
The main goal of Graphalytics is to enable the fair and objective comparison of graph analysis platforms. 
Due to the diversity of bottlenecks and performance issues such platforms need to address, Graphalytics consists of a set of selected deterministic algorithms for full-graph analysis, standard graph datasets, synthetic dataset generators, and reference output for validation purposes. 
Its test harness produces deep metrics that quantify multiple kinds of systems scalability, weak and strong, and robustness, such as failures and performance variability. 
The benchmark also balances comprehensiveness with runtime necessary to obtain the deep metrics. 
The benchmark comes with open-source software for generating performance data, for validating algorithm results, 
for monitoring and sharing performance data, and 
for obtaining the final benchmark result as a standard performance report. 
}

\execSummary{

Processing graphs, especially at large scale, is an increasingly important activity in a variety of business, engineering, and scientific domains. Tens of very different graph processing platforms, such as Giraph, GraphLab, and even generic data processing frameworks such as Hadoop and Spark, can be used for this purpose. For graph processing platforms to be adopted and to continue their evolution, users must be able to select with ease the best-performing graph processing platform, and developers and system integrators have to find it easy to quantify the non-functional aspects of the system, from performance to scalability. Compared to traditional benchmarking, benchmarking graph processing platforms must provide a diverse set of kernels (algorithms), provide recipes for generating diverse yet controlled datasets at large scale, and be portable to diverse and evolving platforms. 

The Linked Data Benchmark Council's Graphalytics is a combined industry and academia initiative, formed by 
principal actors in the field of graph-like data management. 
The main goal of LDBC Graphalytics is to define a benchmarking framework, and the associated 
open-source software tools, where different graph processing platforms and core graph data management
technologies can be fairly tested and compared.
The key feature of Graphalytics is the understanding of the irregular and deep impact that dataset and algorithm diversity can have on performance, leading to bottlenecks and performance issues.
For this feature, Graphalytics proposes a set of selected deterministic algorithms for full-graph analysis, standard graph datasets, synthetic dataset generators, and reference output for validation purposes. 
Furthermore, the Graphalytics test harness can be used to conduct diverse experiments and produce deep metrics that quantify multiple kinds of systems scalability, weak and strong, and robustness, such as failures and performance variability. 
The benchmark also balances comprehensiveness with runtime necessary to obtain the deep metrics. 
Because issues change over time, LDBC Graphalytics also proposes a renewal process. 

Overall, Graphalytics aims to drive not only the selection of adequate graph processing platforms, but also help with system tuning by providing data for system-bottleneck identification, and with feature and even system \mbox{(re-)design} by providing quantitative evidence of the presence of sub-optimal designs.
To this end, the development of the benchmark follows and extends the guidelines set by other LDBC benchmarks, such as the LDBC Social Network Benchmark's Interactive and Business Intelligence workloads. 

To increase adoption by industry and research organizations, 
LDBC Graphalytics provides all the necessary software to run its comprehensive benchmark process.
The open-source software contains tools for generating performance data, for validating algorithm results, 
and or monitoring and sharing performance data. 
The software is designed to be easy to use and deploy at a small cost.
Last, the software is developed using modern software engineering practices, 
with low technical debt, high quality of code, and deep testing and validation processes prior to release. 

This preliminary version of the LDBC Graphalytics specification contains 
a formal definition of the benchmarking framework and of its components, 
a description of the benchmarking process,
links to the open-source software including a working benchmarking harness and drivers for several vendor- and community-driven graph processing platforms,
pseudo-code for all algorithms included in the benchmark,
and a comparison with related tools, products, and concepts in the benchmarking space.

}

\begin{document}

\maketitle

\listoffigures
\listoftables


\chapter{Introduction}
\label{chap:introduction}
In this work we introduce the LDBC Graphalytics benchmark, explaining the motivation for creating this benchmark suite for graph analytics platforms, its relevance to industry and academia, the overview of the benchmark process, and the participation of industry and academia in developing this benchmark. A scientific companion to this technical specification has been published in 2016~\cite{DBLP:journals/pvldb/IosupHNHPMCCSATXNB16}.

\section{Motivation for the Benchmark}
Responding to increasingly larger and more diverse graphs, and the need to analyze them, both industry and academia are developing and tuning graph analysis software platforms. Already tens of such platforms exist,	but their performance is often difficult to compare. Moreover, the random, skewed, and correlated access patterns of graph analysis, caused by the complex interaction between input datasets and applications processing them, expose new bottlenecks on the hardware level, as hinted at by the large differences between Top500 and Graph500 rankings (see \autoref{chap:related-work} for the related work). Therefore, addressing the need for fair, comprehensive, standardized comparison of graph analysis platforms, in this work we propose the LDBC Graphalytics benchmark.

\section{Relevance to Industry and Academia}
A standardized, comprehensive benchmark for graph analytics platforms is beneficial to both industry and academia. Graphalytics allows a comprehensive, fair comparison across graph analysis platforms. The benchmark results provide insightful knowledge to users and developers on performance tuning of graph processing, and increases the understanding of the advantages and disadvantages of the design and implementation, therefore stimulating academic research in graph data storage, indexing, and analysis. By supporting platform variety, it reduces the learning curve of new users to graph processing systems.

\section{General Benchmark Overview}
This benchmark suite evaluates the performance of graph analysis platforms that facilitate complex and holistic graph computations. This benchmark must comply to the following requirements:
(R1)~targeting platforms and systems;
(R2)~incorporating diverse, representative benchmark elements;
(R3)~using a diverse, representative process;
(R4)~including a renewal process;
(R5)~developed under modern software engineering techniques.

In the benchmark (see \autoref{chap:definition} for the formal definition), we carefully motivate the choice of our algorithms and datasets to conduct our benchmark experiments. Graphalytics consists of six core algorithms (also known as kernels~\cite{DBLP:conf/hipc/BaderM05}): breadth-first search, PageRank, weakly connected components, community detection using label propagation, local clustering coefficient, and single-source shortest paths. The workload includes real and synthetic datasets, which are classified into intuitive ``T-shirt'' sizes (e.g., S, M, L, XL). The benchmarking process is made future-proof, through a {\it renewal process}. 

Each system under test undergoes a standard benchmark (see \autoref{chap:benchmark_process}) per target scale, which executes in total 90 graph-processing jobs (six core algorithms, five different datasets, and three repetitions per job).
\futureinversion{2.0}{For full benchmark, additional experiments on scalability and robustness will also be included. Our test harness characterizes performance and {\it scalability} with deep metrics (strong vs.\ weak scaling), and also characterizes {\it robustness} by measuring SLA compliance, performance variability, and crash points.}

\section{Participation of Industry and Academia}
The Linked Data Benchmark Council (\url{ldbcouncil.org}, LDBC), is an industry council formed to 
establish standard benchmark specifications, practices and results for {\em graph data management systems}. The list of institutions that take part in the definition and development of LDBC Graphalytics is formed by relevant actors from both the industry and academia in the field of large-scale graph processing. As of February 2017, the list of participants is as follows:

\begin{itemize}
	\item \MakeUppercase{Centrum Wiskunde \& Informatica}, the Netherlands (CWI)
	\item \MakeUppercase{Delft University of Technology}, the Netherlands (TUD)
	\item \MakeUppercase{Vrije Univrsiteit Amsterdam}, the Netherlands (VU)
	\item \MakeUppercase{Georgia Institute of Technology}, USA (GT)
	\item \MakeUppercase{Huawei Research America}, USA (HUAWEI)
	\item \MakeUppercase{Intel Labs}, USA (INTEL)
	\item \MakeUppercase{Oracle Labs}, USA (ORACLE)
	\item \MakeUppercase{Polytechnic University of Catalonia}, Spain (UPC)
	\item \MakeUppercase{MTA-BME Lend\"ulet Research Group on Cyber-Physical Systems at the Budapest University of Technology and Economics}, Hungary (MTA-BME)
\end{itemize}

\section{Technical report}

This technical report is available on arXiv~\cite{DBLP:journals/corr/abs-2011-15028}.

\chapter{Formal Definition}
\label{chap:definition}

\section{Requirements}
\label{sec:requirements}
The Graphalytics benchmark is the result of a number of design choices.

\begin{itemize}
\item[\textbf{(R1)}] \textbf{Target platforms and systems:} benchmarks must support any graph analysis platform operating on any underlying hardware system. For platforms, we do not distinguish between programming models and support any model. For systems, we target the following environments: 
multi-core and many-core single-node systems, systems with accelerators (GPUs, FPGAs, ASICs), hybrid systems, and distributed systems that possibly combine several of the previous types of environments.
Without R1, a benchmark could not service a diverse industrial following.

\item[\textbf{(R2)}] \textbf{Diverse, representative benchmark elements:} data model and workload selection must be representative and have a good coverage of real-world practice. In particular, the workload selection must include datasets and algorithms which cover known system bottlenecks and be representative in the current and near-future practice. Without representativeness, a benchmark could bias work on platforms and systems towards goals that are simply not useful for improving current practice. Without coverage, a benchmark could push the community into pursuing cases that are currently interesting for industry, but not address what could become impassable bottlenecks in the near-future.

\item[\textbf{(R3)}] \textbf{Diverse, representative process:} the set of experiments conducted by the benchmark automatically must be broad, covering the main bottlenecks of the target systems. \futureinversion{2.0}{In particular, the target systems are known to raise various scalability issues, and also, because of deployment in real-world clusters, be prone to various kinds of failures, exhibit performance variability, and overall have various robustness problems.} The process must also include validation of results, thus making sure the processing is done correctly. Without R3, a benchmark could test very few of the diverse capabilities of the target platforms and systems, and benchmarking results could not be trusted.

\item[\textbf{(R4)}] \textbf{Include renewal process:} unlike many other benchmarks, benchmarks in the area of graph processing must include a renewal process, that is, not only a mechanism to scale up or otherwise change the workload to keep up with increasing more powerful systems, but also a process to automatically configure the mechanism, and a way to characterize the reasonable characteristics of the workload for an average platform running on an average system. Without R4, a benchmark could become less relevant for the systems of the future.

\item[\textbf{(R5)}] \textbf{Modern software engineering:} benchmarks must include a modern software architecture and run a modern software-engineering process. The Graphalytics benchmark is provided with an extensive benchmarking suite that allows users to easily add new platforms and systems to test. This makes it possible for practitioners to easily access the benchmarks and compare their platforms and systems against those of others. Without R5, a benchmark could easily become unmaintainable or unusable.
\end{itemize}

\section{Data}
The Graphalytics benchmark operates on a single type of dataset: graphs. This section provides the definition of a graph, the definition of the representation used by Graphalytics for its input and output data, and the datasets used for the benchmarks.

\subsection{Definition}
Graphalytics does not impose any requirements on the semantics of the graph. The benchmark uses a typical data model for graphs. A graph consists of a collection of \emph{vertices} (nodes) which are linked by \emph{edges} (relationships).  Each vertex is assigned a unique identifier represented as an unsigned 64-bit integer, i.e., between $0$ and $2^{64}-1$. Vertex identifiers do not necessarily start at zero, nor are the identifiers necessarily consecutive. Edges are represented as pairs of vertex identifiers. Graphalytics supports both \emph{directed} graphs (i.e., edges are unidirectional) and \emph{undirected} graphs (i.e., edges are bidirectional). Every edge is unique and connects two distinct vertices. This implies that self-loops (i.e., vertices having edges to themselves) and multi-edges (i.e., multiple edges between a pair of vertices) are not allowed. In the case of undirected graphs, for every pair of vertices $u$ and $v$, the edges $(u,v)$ and $(v,u)$ are considered to be identical.

Vertices and edges can have properties which can be used to store meta-data such as weights, timestamps, labels, costs, or durations. Currently, Graphalytics supports three types of properties: \emph{integers}, \emph{floating-point numbers}, and \emph{booleans}. All floating-point numbers must be internally stored and handled in 64-bit double-precision IEEE 754 format. We explicitly do not allow the single precision format, as this can speedup computation and presents an unfair advantage. The current specification of Graphalytics does not use integer or boolean properties.

\subsection{Representation}\label{sec:data:representation}
In Graphalytics, the file format used to represent the graphs is the ``Edge/Vertex-List with Properties'' (\emph{EVLP}) format for graphs. The format consists of two text files: a vertex-file containing the vertices (with optional properties) and an edge-list containing the edges (with optional properties). Both files are plain text (ASCII) and consist of a sequence of lines.

For the vertex files, each line contains exactly one vertex identifier. The vertex identifiers are sorted in ascending order to facilitate easy conversion to other formats. For the edge files, each line contains two vertex identifiers separated by a space. The edges are sorted in lexicographical order based on the two vertices. For directed graphs, the source vertex is listed first. For undirected graphs, the smallest identifier of the two vertices is listed first and each edge is only listed once in one direction.

Vertices and edges can have optional properties. These values of these properties are listed in the vertex/edge files after each vertex/edge and are separated by spaces. The interpretation of these properties is not provided by the files.

\begin{figure}[t!]
\centering
\begin{subfigure}{0.3\textwidth}
\includegraphics[width=0.9\textwidth]{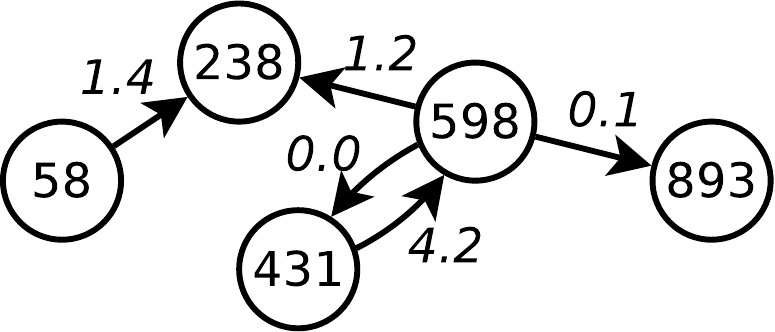}
\end{subfigure}
\begin{subfigure}{0.3\textwidth}
\begin{Verbatim}[frame=single]
58 
238
431
598
893
\end{Verbatim}
\end{subfigure}
\begin{subfigure}{0.3\textwidth}
\begin{Verbatim}[frame=single]
58 238 1.4
431 598 4.2
598 238 1.2
598 431 0.0
598 893 0.1
\end{Verbatim}
\end{subfigure}
\newline
\newline
\begin{subfigure}{0.3\textwidth}
  \centering
  (a) Graph
\end{subfigure}
\begin{subfigure}{0.3\textwidth}
  \centering
  (b) Vertex file
\end{subfigure}
\begin{subfigure}{0.3\textwidth}
  \centering
  (c) Edge file
\end{subfigure}

\caption{Example of directed weighted graph in EVLP format.}
\label{fig:definition_example_evlp}
\end{figure}

\autoref{fig:definition_example_evlp} shows an example of the EVLP format for a small directed weighted graph consisting of 5 vertices and 5 directed edges.

\subsection{Size and Scale}
\label{sec:definition_scale}

Graphalytics includes both graphs from real-world applications and synthetic graphs which are created using graph generators. Graphalytics uses a broad range of graphs with a large variety in domain, size, density, and other characteristics. To facilitate performance comparison across datasets, the \emph{scale} of a graph is derived from the number of vertices ($n$) and the number of edges ($m$). Formally, the scale of a graph is defined by calculating the sum $n + m$, taking the logarithm of this value in base 10, and truncating the result to one decimal place. Formally, this can be written as follows:
\begin{equation}
\textit{Scale}(n, m) = \lfloor 10 \log_{10}(n + m) \rfloor / 10
\end{equation}
The scale of a graph gives users an intuition of what the size of graph means in practice. Scales are grouped into classes spanning 0.5 \emph{scale units} and these classes are labeled using familiar system of ``T-shirt sizes'': small (S), medium (M), and large (L), with extra (X) prepended for extremes scales. The reference point is class L, which is defined by the Graphalytics team as the largest class such that the BFS algorithm completes within an hour on any graph from that scale using a state-of-the-art graph analysis platform and a single commodity machine. \autoref{tab:definition_scales} summarizes the classes used in Graphalytics.

\begin{table}
\centering
\begin{tabular}{|l||c|}
\hline
\textbf{Label} & \textbf{Scales} \\ 
\hline \hline
2XS & $6.5 - 6.9$ 
\\ \hline
XS & $7.0 - 7.4$ 
\\ \hline
S & $7.5-7.9$ 
 \\ \hline
M & $8.0 - 8.4$ 
\\ \hline
L & $8.5 - 8.9$ 
\\ \hline
XL & $9.0 - 9.4$ 
 \\ \hline
2XL & $9.5 - 9.9$ 
\\ \hline
3XL & $10.0 - 10.4$ 
\\ \hline
\end{tabular}
\caption{Mapping of dataset scales (``T-shirt sizes'') in Graphalytics.}
\label{tab:definition_scales}
\end{table}

\subsection{Datasets}\label{sec:definition_datasets}
Graphalytics uses both graphs from real-world applications and synthetic graphs which are generated using data generators, the selection of which spans a variety of sizes and densities.

The Graphalytics data sets are available in the SURF/CWI data repository~\cite{cwi:graphalytics} at \url{https://repository.surfsara.nl/datasets/cwi/graphalytics}.

\paragraph{Real-world Datasets} By including real-world graphs from a variety of domains, Graphalytics covers users from different communities, including graphs from the knowledge, gaming, and social network domains.  \autoref{tab:real-datasets} lists the real-world datasets used by the Graphalytics benchmark.

\begin{table}[h]
\centering
\begin{tabular}{|l|l|r|r|r|l|}
\hline
\textbf{ID} & \textbf{Name} & \textbf{$n$} & \textbf{$m$} & \textbf{Scale} & \textbf{Domain} \\
\hline
R1(2XS) & wiki-talk~\cite{snapnets} & 2.39\,M & 5.02\,M & 6.9 & Knowledge \\
\hline
R2(XS) & kgs~\cite{DBLP:conf/netgames/GuoI12} & 0.83\,M & 17.9\,M & 7.3 & Gaming \\
\hline
R3(XS) & cit-patents~\cite{snapnets} & 3.77\,M & 16.5\,M & 7.3 & Knowledge \\
\hline
R4(S) & dota-league~\cite{DBLP:conf/netgames/GuoI12} & 0.06\,M & 50.9\,M & 7.7 & Gaming \\
\hline
R5(XL) & com-friendster~\cite{snapnets} & 65.6\,M & 1.81\,B & 9.3 & Social \\
\hline
R6(XL) & twitter\_mpi~\cite{DBLP:conf/icwsm/ChaHBG10} & 52.6\,M & 1.97\,B & 9.3 & Social \\
\hline
\end{tabular}
\caption{Real-world datasets used by Graphalytics.}
\label{tab:real-datasets}
\end{table}

\futureinversion{2.0}{Renew the graph selection process for real-world graphs, using new graph data archive, for example, with suggestion from Mihai....}

\paragraph{Synthetic Datasets} Besides real-world datasets, Graphalytics adopts two commonly used generators that generate two types of graphs: power-law graphs from the \textbf{Graph500} generator~\cite{chakrabarti2004, murphy2010} and social network graphs from \textbf{LDBC Datagen}~\cite{DBLP:conf/sigmod/ErlingALCGPPB15}. Tables~\ref{tab:graph500-datasets} and \ref{tab:datagen-datasets} list the Graph500 datasets and the Datagen datasets.

\begin{table}[h]
\centering
\begin{tabular}{|l|l|r|r|r|}
	\hline
	\textbf{ID} & \textbf{Name} &   $n$ &    $m$ & \textbf{Scale} \\ \hline\hline
	G22(S)      & Graph500-22   &  2.4M &  64.2M &            7.8 \\ \hline
	G23(M)      & Graph500-23   &  4.6M & 129.3M &            8.1 \\ \hline
	G24(M)      & Graph500-24   &  8.9M & 260.4M &            8.4 \\ \hline
	G25(L)      & Graph500-25   & 17.0M & 523.6M &            8.7 \\ \hline
	G26(XL)     & Graph500-26   & 32.8M &   1.1B &            9.0 \\ \hline
	G27(XL)     & Graph500-27   & 65.6M &   2.1B &            9.3 \\ \hline
	G28(2XL)    & Graph500-28   &  121M &   4.2B &            9.6 \\ \hline
	G29(2XL)    & Graph500-29   &  233M &   8.5B &            9.9 \\ \hline
	G30(3XL)    & Graph500-30   &  448M &  17.0B &           10.2 \\ \hline
\end{tabular}
\caption{Synthetic Graph500 datasets used by Graphalytics.}
\label{tab:graph500-datasets}
\end{table}

\begin{table}[t!]
\centering
\begin{tabular}{|l|l|r|r|r|}
\hline
\textbf{ID} & \textbf{Name} & $n$ & $m$ & \textbf{Scale} \\
\hline \hline
D7.5(S) & Datagen-7.5-fb & 0.6M & 34.2M & 7.5 \\ \hline
D7.6(S) & Datagen-7.6-fb & 0.8M & 42.2M & 7.6 \\ \hline
D7.7(S) & Datagen-7.7-zf & 13.2M & 32.8M & 7.6 \\ \hline
D7.8(S) & Datagen-7.8-zf & 16.5M & 41.0M & 7.7 \\ \hline
D7.9(S) & Datagen-7.9-fb & 1.4M & 85.7M & 7.9 \\ \hline
D8.0(M) & Datagen-8.0-fb & 1.7M & 107.5M & 8.0 \\ \hline
D8.1(M) & Datagen-8.1-fb & 2.1M & 134.3M & 8.1 \\ \hline
D8.2(M) & Datagen-8.2-zf & 43.7M & 106.4M & 8.1 \\ \hline
D8.3(M) & Datagen-8.3-zf & 53.5M & 130.6M & 8.2 \\ \hline
D8.4(M) & Datagen-8.4-fb & 3.8M & 269.5M & 8.4 \\ \hline
D8.5(L) & Datagen-8.5-fb & 4.6M & 332.0M & 8.5 \\ \hline
D8.6(L) & Datagen-8.6-fb & 5.7M & 422.0M & 8.6 \\ \hline
D8.7(L) & Datagen-8.7-zf & 145.1M & 340.2M & 8.6 \\ \hline
D8.8(L) & Datagen-8.8-zf & 168.3M & 413.4M & 8.7 \\ \hline
D8.9(L) & Datagen-8.9-fb & 10.6M & 848.7M & 8.9 \\ \hline
D9.0(XL) & Datagen-9.0-fb & 12.9M & 1.0B & 9.0 \\ \hline
D9.1(XL) & Datagen-9.1-fb & 16.1M & 1.3B & 9.1 \\ \hline
D9.2(XL) & Datagen-9.2-zf & 434.9M & 1.0B & 9.1 \\ \hline
D9.3(XL) & Datagen-9.3-zf & 555.3M & 13.1B & 9.2 \\ \hline
D9.4(XL) & Datagen-9.4-fb & 29.3M & 2.6B & 9.4 \\ \hline
D-3k(XL) & Datagen-sf3k-fb & 33.5M & 2.9B & 9.4 \\ \hline
D-10k(2XL) & Datagen-sf10k-fb & 100.2M & 9.4B & 9.9 \\ \hline
\end{tabular}
\caption{Synthetic Datagen datasets used by Graphalytics.}
\label{tab:datagen-datasets}
\end{table}

\section{Algorithms}
\label{sec:definition_algorithms}

The Graphalytics benchmark consists of six algorithms (also known as \emph{kernels}~\cite{DBLP:conf/hipc/BaderM05}) which need to be executed on the different datasets: five algorithms for unweighted graphs and one algorithm for weighted graphs. These algorithms have been selected based on the results of multiple surveys and expert advice from the participants of the LDBC Technical User Community (TUC) meeting.

Each workload of Graphalytics consists of executing a single algorithm on a single dataset. Below, abstract descriptions are provided for the six algorithms; pseudo-code is given in \autoref{chap:algorithms}. Furthermore, a link to the reference implementation is presented in \autoref{sec:instructions:core}. However, Graphalytics does not impose any constraint on the implementation of algorithms. Any implementation is allowed, as long as its correctness can be validated by comparing its output to correct reference output (\autoref{sec:definitions_validation}).

In the following sections, a graph $G$ consists of a set of vertices $V$ and a set of edges $E$. For undirected graphs, each edge is bidirectional, so if $(u,v)\in E$ then $(v,u)\in E$. Each vertex $v$ has a set of outgoing neighbors
$N_\mathrm{out}(v) = \{u \in V | (v, u) \in E \}$, a set of incoming neighbors
$N_\mathrm{in}(v)  = \{u \in V | (u, v) \in E \}$, and a set of all neighbors
$N(v) = N_\mathrm{in}(v) \cup N_\mathrm{out}(u)$.
Note that for undirected graphs, each edge is bidirectional so we have $N_\mathrm{in}(v) = N_\mathrm{out}(v) = N(v)$.

\subsection{Breadth-First Search (BFS)}
\label{sec:bfs}
\emph{Breadth-First Search} is a traversal algorithm that labels each vertex of a graph with the length (or \emph{depth}) of the shortest path from a given source vertex (\emph{root}) to this vertex. The root has depth $0$, its outgoing neighbors have depth $1$, their outgoing neighbors have depth $2$, etc. Unreachable vertices should be given the value infinity
(represented as \texttt{9223372036854775807}).
Example graphs are shown in \autoref{fig:bfs_example}.

\subsection{PageRank (PR)}
\label{sec:pr}
\emph{PageRank} is an iterative algorithm that assigns to each vertex a ranking value. The algorithm was originally used by Google Search to rank websites in their search results~\cite{page1999pagerank}. Let $\textit{PR}_i(v)$ be the PageRank value of vertex $v$ after iteration $i$. Initially, each vertex $v$ is assigned the same value such that the sum of all vertex values is $1$.
\begin{equation}
\textit{PR}_0(v) = \frac{1}{|V|}
\end{equation}
After iteration $i$, each vertex pushes its PageRank over its outgoing edges to its neighbors. The PageRank for each vertex is updated according to the following rule:
\begin{equation}
\textit{PR}_i(v) =
  \underbrace{\frac{1-d}{|V|}}_\textit{teleport} +
  \underbrace{d \cdot \sum_{u \in N_\mathrm{in}(v)} \frac{\textit{PR}_{i - 1}(u)}{|N_\mathrm{out}(u)|}}_\textit{importance} +
  \underbrace{\frac{d}{|V|} \cdot \sum_{w \in D} \textit{PR}_{i - 1}(w)}_\textit{redistributed from sinks}
\end{equation}
Notation: $d \in [0,1]$ is the \emph{damping factor} and $D = \big\{w \in V \ \big| \ |N_\mathrm{out}(w)| = 0 \big\}$ is the set of \emph{sink vertices}, i.e., vertices having no outgoing edges.
Sink vertices have nowhere to push their PageRank to, so the total sum of the PageRanks for the sink vertices is evenly distributed over all vertices~\cite{DBLP:conf/www/EironMT04}.
When computing the $\textit{importance}$ value, we consider $\frac{\textit{PR}_{i-1}(u)}{|N_\mathrm{out}(u)|}$ to be $0$ for sink vertices. 

The PageRank algorithm should continue for a fixed number of iterations. The floating-point values must be handled as 64-bit double-precision IEEE 754 floating-point numbers.

\subsection{Weakly Connected Components (WCC)}
\label{sec:wcc}
This algorithm finds the \emph{Weakly Connected Components} of a graph and assigns each vertex a unique label that indicates which component it belongs to. Two vertices belong to the same component, and thus have the same label, if there exists a path between these vertices along the edges of the graph. For directed graphs, it is allowed to travel over the reverse direction of an edge, i.e., the graph is interpreted as if it is undirected.
Example graphs are shown in \autoref{fig:wcc_example}.

\subsection{Community Detection using Label Propagation (CDLP)}
\label{sec:cdlp}
The \emph{community detection} algorithm in Graphalytics uses \emph{label propagation} (CDLP) and is based on the algorithm proposed by Raghavan et al.~\cite{raghavan2007near}. The algorithm assigns each vertex a label, indicating its community, and these labels are iteratively updated where each vertex is assigned a new label based on the frequency of the labels of its neighbors. The original algorithm has been adapted to be both deterministic and parallel, thus enabling output validation and parallel execution.

Let $L_i(v)$ be the label of vertex $v$ after iteration $i$. Initially, each vertex $v$ is assigned a unique label which matches its identifier.
\begin{equation}
L_0(v) = v
\end{equation}
In iteration $i$, each vertex $v$ determines the frequency of the labels of its incoming and outgoing neighbors and selects the label which is most common.  If the graph is directed and a neighbor is reachable via both an incoming and outgoing edge, its label will be counted twice. In case there are multiple labels with the maximum frequency, the smallest label is chosen. In case a vertex has no neighbors, it retains its current label. This rule can be written as follows:
\begin{equation}
L_i(v) = \min \left( \underset{l}{\mathrm{arg\,max}} \Bigg[ \Big|\{ u \in N_\mathrm{in}(v)~|~L_{i-1}(u) = l \}\Big| + \Big|\{ u \in N_\mathrm{out}(v)~|~L_{i-1}(u) = l \}\Big| \Bigg] \right)
\end{equation}

Example graphs are shown in \autoref{fig:cdlp_example}.

\emph{Note.} The CDLP algorithm in \toolname{} has two key differences from the original algorithm proposed in~\cite{raghavan2007near}.
First, it is deterministic: if there are multiple labels with their frequency equalling the maximum, it selects the smallest one while the original algorithm selects randomly.
Second, it is synchronous, i.e., each iteration is computed based on the labels obtained as a result of the previous iteration.
As remarked in~\cite{raghavan2007near}, this can cause the oscillation of labels in bipartite or nearly bipartite subgraphs.

\subsection{Local Clustering Coefficient (LCC)}
\label{sec:lcc}
The \emph{Local Clustering Coefficient} algorithm determines the local clustering coefficient for each vertex. This coefficient indicates the ratio between the number of edges between the neighbors of a vertex and the maximum number of possible edges between the neighbors of this vertex. If the number of neighbors of a vertex is less than two, its coefficient is defined as zero. The definition of LCC can be written as follows:
\begin{equation}
LCC(v) = \begin{cases}
0 & \text{If } |N(v)| \leq 1 \\
\frac{|\{(u, w) | u, w \in N(v) \wedge (u, w) \in E\}|}
{|{(u,w) | u, w \in N(v)}|} & \text{Otherwise} \\
\end{cases}
\end{equation}
Note that the second case can also be written using the sum over the neighbors of $v$.
\begin{equation}
LCC(v) = \frac{\sum_{u \in N(v)} |N(v) \cap N_\mathrm{out}(u)|}{|N(v)| \big( |N(v)| - 1 \big)}
\end{equation}
For \emph{directed graphs}, the set of neighbors $N(v)$ is determined without taking directions into account, but each neighbor is only counted once (a neighbor with both an incoming and an outgoing edge from vertex $v$ does not count twice).
However, directions are enforced when determining $N_\mathrm{out}(v)$ between neighbors. Note that calculating the intersection using the \emph{incoming} edges to $u$ yields the same result, i.e.\
\begin{equation}
	\sum_{u \in N(v)} |N(v) \cap N_\mathrm{out}(u)| = 
	\sum_{u \in N(v)} |N(v) \cap N_\mathrm{in}(u)|
\end{equation}

Example graphs are shown in \autoref{fig:lcc_example} and \autoref{fig:lcc_dir_example_detailed}.

\subsection{Single-Source Shortest Paths (SSSP)}
\label{sec:sssp}
The \emph{Single-Source Shortest Paths} algorithm marks each vertex with the length of the shortest path from a given \emph{root} vertex to every other vertex in the graph. The length of a path is defined as the sum of the weights on the edges of the path. The edge weights are floating-point numbers which must be handled as 64-bit double-precision IEEE 754 floating-point numbers. The edge weights are never negative, infinity, or invalid (i.e., \emph{NaN}), but are allowed to be zero. Unreachable vertices should be given the value infinity (represented as \texttt{infinity}).

Example graphs are shown in \autoref{fig:sssp_example}.

\section{Output Validation}
\label{sec:definitions_validation}

The output of every execution of an algorithm on a dataset must be validated for the result to be admissible. All algorithms in the Graphalytics benchmark are deterministic and can therefore be validated by comparing to reference output for correctness. The reference output is typically generated by a specifically chosen reference platform, the implementation of which is cross-validated with at least two other platforms up to target scale~L. \futureinversion{2.0}{Define target scale in this chapter [Gabor]} The results are tested by cross-validating multiple platforms and implementations against each other.

The validation output presents numbers either as integers or floating-point numbers, depending on the algorithm definition. Note that these numbers are stored in the file system as decimal values in plain text (ASCII).
For floating-point numbers, a scientific notation with 15~significant digits (e.g., $2.476\,533\,217\,845\,853\mathrm{e-}08$) is used.

The system's output generated during the benchmark must be stored as decimal values in plain text (ASCII) files, grouped into a single file directory. The formatting rules for integers must be exactly the same as the validation output.

There are three methods used for validation:

\begin{figure}[h]
\centering
\begin{subfigure}{0.3\textwidth}
\begin{Verbatim}[frame=single]
1 3
2 1
3 2
4 0
5 1
\end{Verbatim}
\caption{Reference output}
\end{subfigure}
\begin{subfigure}{0.3\textwidth}
\begin{Verbatim}[frame=single]
1 3
2 1
3 2
4 0
5 1
\end{Verbatim}
\caption{Example of correct result}
\end{subfigure}
\begin{subfigure}{0.3\textwidth}
\begin{Verbatim}[frame=single,commandchars=\\\{\}]
1 \color{red}4
2 1
3 2
4 \color{red}5
5 1
\end{Verbatim}
\caption{Example of incorrect result}
\end{subfigure}
\caption{Example of validation with \emph{exact match}.}
\label{fig:definition_validation_exact}
\end{figure}

\begin{figure}[h]
\centering
\begin{subfigure}{0.3\textwidth}
\begin{Verbatim}[frame=single]
1 1
2 1
3 1
4 2
5 2
6 3
\end{Verbatim}
\caption{Reference output}
\end{subfigure}
\begin{subfigure}{0.3\textwidth}
\begin{Verbatim}[frame=single]
1 81
2 81
3 81
4 32
5 32
6 12
\end{Verbatim}
\caption{Example of correct result}
\end{subfigure}
\begin{subfigure}{0.3\textwidth}
\begin{Verbatim}[frame=single,commandchars=\\\{\}]
1 31
2 \color{red}52
3 31
4 \color{red}31
5 \color{red}31
6 74
\end{Verbatim}
\caption{Example of incorrect result}
\end{subfigure}
\caption{Example of validation with \emph{equivalence match}.}
\label{fig:definition_validation_equivalence}
\end{figure}

\begin{figure}[h]
\centering
\begin{subfigure}{0.3\textwidth}
\begin{Verbatim}[frame=single]
1 0
2 0.3
3 0.45
4 0.23
5 9223372036854775807
6 0.001
\end{Verbatim}
\caption{Reference output}
\end{subfigure}
\begin{subfigure}{0.3\textwidth}
\begin{Verbatim}[frame=single]
1 0
2 0.30002
3 0.45
4 0.229997
5 9223372036854775807
6 0.001
\end{Verbatim}
\caption{Example of correct result}
\end{subfigure}
\begin{subfigure}{0.3\textwidth}
\begin{Verbatim}[frame=single,commandchars=\\\{\}]
1 \color{red}0.000001
2 0.3
3 \color{red}0.46
4 \color{red}0.22
5 \color{red}1.79769e+308
6 \color{red}0
\end{Verbatim}
\caption{Example of incorrect result}
\end{subfigure}
\caption{Example of validation with \emph{epsilon match}.}
\label{fig:definition_validation_epsilon}
\end{figure}

\begin{itemize}

\item \textbf{Exact match (applies to BFS, CDLP):} the vertex values of the system's output should be identical to the reference output. \autoref{fig:definition_validation_exact} shows an example of validation with exact match.

\item \textbf{Equivalence match (applies to WCC):} the vertex values of the system's output should be equal to the reference output \emph{under equivalence}. This means a two-way mapping should exists that maps the system's output to be identical to reference output and the inverse of this mapping maps the reference output to be identical to the system's output. In other words, the output is considered to be valid if all vertices which have the same label in the system's output also have the same label in the reference output, and vice versa. \autoref{fig:definition_validation_equivalence} shows an example of validation with equivalence.

\item \textbf{Epsilon match (applies to PR, LCC, SSSP):} a margin of error is allowed for some algorithms due to floating-point rounding errors. Let $r$ be the reference value of a vertex and $s$ be the system's output value of the same vertex. These values are considered to match if $s$ is within $0.01\%$ of $r$, i.e., the equation $|r-s| \leq \varepsilon |r|$ holds where $\varepsilon=0.0001$ (equality is allowed such that the case where $r = 0$ and $s = 0$ passes). The value of $\varepsilon$ was chosen such that errors that result from rounding are not penalized. \autoref{fig:definition_validation_epsilon} shows an example of validation with epsilon match.
\end{itemize}

Small validation example graphs are available in the Graphalytics suite and are listed in \autoref{chap:validation_examples}, including a common validation graph for all six algorithms (\autoref{fig:common_example}).

\futureinversion{2.0}{A future validation component of Graphalytics will focus on performance. {\it Why?} The current approach is correct, but slow. Validating XL graphs is already too time consuming and memory-intensive. {\it How?}}

\section{Job}
\label{sec:def:job}
A graph-processing job is the process of executing a graph algorithm (see \autoref{sec:definition_algorithms}) on a graph dataset (see \autoref{sec:definition_datasets}). This section discusses the description of a graph-processing job, the underlying operations constituting a graph processing job, and the metrics used to measure the job performance.

\subsection{Description}
Each graph-processing job is specified by a list of descriptive information, specifically, the system description, the algorithm, the dataset, and the benchmark configuration. The job description should be uniformly applicable to any graph-processing system.

\begin{itemize}
    \item \textbf{System}: platform type, environment type, and system cost.
    \item \textbf{Algorithm}: algorithm type, and algorithm parameters. 
    \item \textbf{Dataset}: vertex size, edge size, and graph size (vertex size + edge size).
    \item \textbf{Benchmark}: deployment mode, allocated resources, and time-out duration.
\end{itemize}

\subsection{Operations}
\label{sec:def:job:operation}
Graph processing is data-intensive, sensitive to the data irregularity, and often involves iterative processing. Typically, a graph-processing system facilitates a \textbf{Loading} phase to pre-process the data, and follows with one or more \textbf{Running} phases to run various graph algorithms on the pre-processed data.

\begin{figure}[h]
	\centering
	\includegraphics[width=0.9\linewidth]{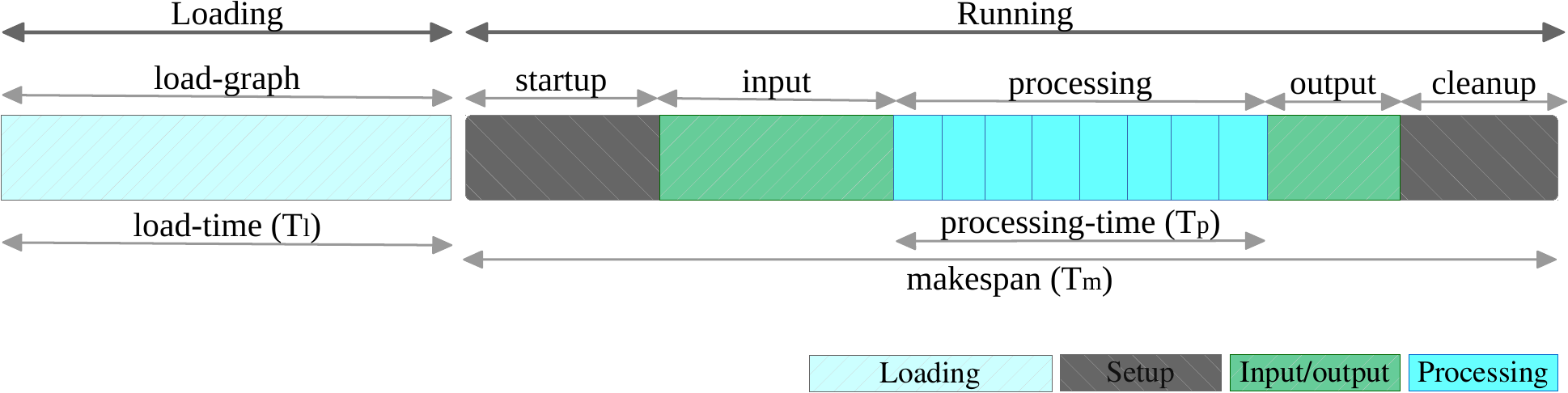}
	\caption{A typical graph processing job with the underlying operations.}
	\label{fig:job}
\end{figure}

During the \textbf{Loading} step, the input graph data can be converted into an optimized system-specific data format and pre-loaded into a local/share/distributed storage system. Binary formats are allowed, however, the pre-processed data must be uniformly usable for any graph algorithms targeting the same graph type (e.g., algorithms on unweighted graphs must use the same binary format).

During the \textbf{Running} step, the graph-processing system carries out a series of operations to facilitate efficient algorithm execution on the pre-processed data. These operations are categorized into three types: setup, input/output, and processing operations (see \autoref{fig:job}).
\begin{itemize}
    \item \textbf{Setup operations} reserve computational resources in distributed environments, prepare the system for operation, clean up the environment after the termination of the job.
    \item \textbf{Input/output operations} transfer graph data from storage to the memory space, and convert the data to specific formats before/after data processing, and offload the outputs back to the storage. Some platforms load/unload data from a distributed file system, other distributed platforms from share/local storage in each node. \futureinversion{2.0}{Discuss that in-memory systems load the graph in this phase [Gabor]}
    \item \textbf{Processing operations} take in-memory data and process it according to an user-defined algorithm and its expression in a programming paradigm. Processing operations typically include iterative ``processing'' steps.
\end{itemize}

\subsection{Metrics} 
\label{sec:def:metrics}
This section describes the metrics used in Graphalytics. The Graphalytics benchmark includes several metrics to quantify the performance and other characteristics of the system under test. The performance of graph analytics systems is measured by the time spent on several phases in the execution of the benchmark. Graphalytics reports performance, throughput metrics, cost metrics, and ratio metrics, as follows.

The {\bf Performance metrics} report the execution time of various platform operations.

\begin{itemize}
	\item {\bf Load time} ($T_l$), in $\textit{seconds}$:  The time spent loading a particular graph into the system under test, including any preprocessing to convert the input graph to a format suitable for the system. This phase is executed once per graph before any commands are issued to execute specific algorithms.
	\item {\bf Makespan:} ($T_m$), in $\textit{seconds}$: The time between the Graphalytics driver issuing the command to execute an algorithm on a (previously uploaded) graph and the output of the algorithm being made available to the driver. The makespan can be further divided into processing time and overhead\futureinversion{2.0}{Define overhead [Gabor]}. The makespan metric corresponds to the operation of a {\it cold graph-processing system}, which depicts the situation where the system is started up, processes a single dataset using a single algorithm, and then is shut down.
	\item {\bf Processing time} ($T_p$), in $\textit{seconds}$: Time required to execute an actual algorithm. This does not include platform-specific overhead, such as allocating resources, loading the graph from the file system, or graph partitioning. The processing time metric corresponds to the operation of an in-production, {\it warmed-up graph-processing system}, where especially loading of the graph from the file system and graph partitioning, both of which are typically done only once and are algorithm-independent, are not considered.
\end{itemize}

The execution time is capped by the time-out duration configured in each benchmark. Once the time-out is reached, the graph-processing job is terminated, and the time-out duration is reported as the performance metrics instead.

The {\bf Throughput metrics} focus on the processing rate of the system under test. They use a notion of workload intensity, expressed in the graph-specific number of processed edges and vertices:
\begin{itemize}
	\item {\bf Edges Per Second} ($\textit{EPS}$), in $\textit{units} / \textit{second}$: The ratio between the number of edges processed (edge size) and the processing time ($T_p$) is used by other benchmarks (Graph500 in \autoref{tab:SummaryOfRelatedWorkAcronyms}) to quantify the rate of operation of the system under test. This is fine for edge-dominated algorithms, such as the BFS used in the same benchmarks, but does not explain the performance of vertex-dominated algorithms or of algorithms whose performance is a complex function of the structural properties of the dataset.
	
	\item {\bf Edges and Vertices per Second} ($\textit{EVPS}$), in $\textit{units} / \textit{second}$: Graphalytics uses the ratio between the sum of the number of edges and the number of vertices (graph size) processed by the system, and the processing time ($T_p$). EVPS is closely related to the scale of a graph, as defined by Graphalytics (see \autoref{sec:definition_scale}).
\end{itemize}

Graphalytics also reports {\bf Cost metrics}:

\begin{itemize}
	\item {\bf Three-year Total Cost of Ownership} ($\textit{TCO}$), in $\textit{dollars}$: reported in compliance with the LDBC rules~\cite{ldbc_byelaws}, so {\it not} computed by Graphalytics. In particular, LDBC currently adapts the TPC standard pricing model v2.0.0~\cite{tpc_pricing}.
	
	\item {\bf Price-per-performance} ($\textit{PPP}$), in $\textit{dollars} / \textit{unit}$: as the ratio between TCO and EVPS. This is a metric included for compliance with the LDBC charter~\cite{ldbc_byelaws}.

	\futureinversion{2.0}{\item {\bf Three-year Total Energy Costs} ($\textit{TEC}$), in $\textit{watts}$: reported in compliance with the SPEC Power, so {\it not} computed by Graphalytics and {\it not} relying only on the metrics provided by the operating system software and hardware performance counters of current processors/systems.}
	
	\futureinversion{2.0}{\item {\bf Energy-per-performance} ($\textit{EPP}$), in $\textit{watts} / \textit{dollar}$: as the ratio between TEC and EVPS. This is a metric included for compliance with the LDBC charter~\cite{ldbc_byelaws}.}

\end{itemize}

\futureinversion{2.0}{
The {\bf Ratio metrics} reported by Graphalytics are:

\begin{itemize}
	\item {\bf Speedup}: The ratio between processing times for scaled and baseline resources. This metric is used to quantify the scalability of the system under test. 
	The baseline is defined as the performance achieved by a reference graph-processing platform defined by the Graphalytics team until the next revision (here, the public version of Giraph with the reference drivers), in a reference environment defined by the Graphalytics team until the next revision (here, DAS5 in the Netherlands). The renewal process of Graphalytics (see \autoref{sec:renewal}) can change the baseline.
\end{itemize}
}

\section{System Under Test} \label{sec:sut}

Responding to requirement R1 (see \autoref{sec:requirements}), the LDBC Graphalytics framework defines the System Under Test as the combined software platform and hardware environment that is able to execute graph-processing algorithms on graph datasets. This is an inclusive definition, and indeed Graphalytics has been executed in the lab of SUTs with software ranging from community-driven prototype systems, to vendor-optimized software; and with hardware ranging from beefy single-node multi-core systems, to single-node CPU and (multi-)GPU hybrid systems, to multi-node clusters with or without GPUs present. 


\section{Renewal Process} \label{sec:renewal}

To ensure the relevance of Graphalytics as a benchmark for future graph analytics systems, a renewal process is included in Graphalytics. This renewal process updates the workload of the benchmark to keep it relevant for increasingly powerful systems and developments in the graph analytics community. This results in a benchmark which is future-proof. Renewing the benchmark means renewing the algorithms as well as the datasets. For every new version of Graphalytics, a two-stage selection process will be used by the LDBC Graphalytics Task Force. The same selection process was used to derive the workload in the current version of the Graphalytics benchmark.

To achieve both workload representativeness and workload coverage, a two-stage selection process is used. \futureinversion{2.0}{This piece of text is repetitive [Gabor]} The first stage identifies classes of algorithms and datasets that are representative for real-world usage of graph analytics systems. In the second stage, algorithms and datasets are selected from the most common classes such that the resulting selection is diverse, i.e., the algorithms cover a variety of computation and communication patterns, and the datasets cover a range of sizes and a variety of graph characteristics.

Updated versions of the Graphalytics benchmark will also include renewed definitions of the scale classes defined in \autoref{sec:definition_scale}. The definition of the scale classes is derived from the capabilities of state-of-the-art graph analytics systems and common-off-the-shelf machines, which are expected to be improved over time. Thus, graphs that are considered to be large as of the publication of the first edition of Graphalytics (labeled $L'16$ to indicate the $2016$ edition) may be considered medium-sized graphs in the next edition (e.g., $M'20$).

\chapter{Benchmark Process}
\label{chap:benchmark_process}
The Graphalytics benchmark suite is developed to facilitate the benchmark process described in this technical specification. This chapter describes the benchmark composition, the benchmark type, the detailed steps of the benchmark execution, and the format of the benchmark report.

\section{Benchmark}
\label{sec:process:benchmark}
\begin{figure}[h]
	\centering
	\includegraphics[width=0.6\linewidth]{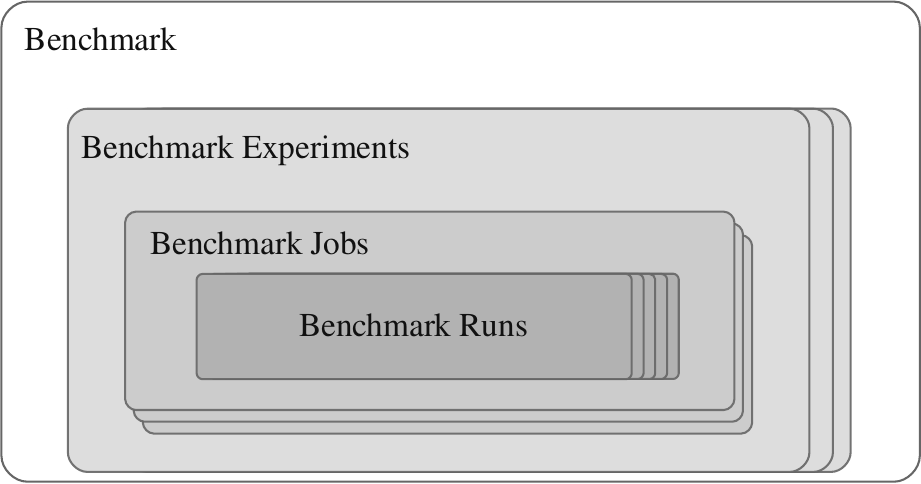}
	\caption{The composition of a benchmark.}
	\label{fig:benchmark_composition}
\end{figure}

A benchmark is a standardized process to quantify the performance of the system under test. \autoref{fig:benchmark_composition} depicts the benchmark composition: each benchmark contains a set of benchmark experiments, each experiment consists of multiple benchmark jobs, and each job is executed repeatedly in the form of benchmark runs.

\begin{itemize}
    \item A \textbf{benchmark experiment} addresses a specific performance characteristic of the system under test, e.g., the performance of an algorithm, or the weak scalability of a system. Each experiment gathers benchmark results from multiple benchmark jobs to quantify a specific performance characteristic.
    
    \item A \textbf{benchmark job} describes, uniformly across all system under tests, the exact job specification of a graph-processing job (see \autoref{sec:def:job}). The job specification contains information, e.g., the system under test (the platform and the environment), the type of algorithm and dataset, and how much resources are used. These information instructs how the system should be configured during the benchmark execution.
    
    \item A \textbf{benchmark run} is a real-world execution of a benchmark job. To gather statistically reliable benchmark results, each benchmark job is repeated multiple times in the form of a benchmark run to mitigate the performance variability during the benchmark execution. 
\end{itemize}

\section{Benchmark Type}
\label{sec:process:type}
The Graphalytics benchmark suite supports four types of benchmark: $\textit{test}$, $\textit{standard}$, $\textit{full}$, and $\textit{custom}$. This section describes the differences and the composition of these four benchmark types.




\subsection{Competition Benchmark}
\label{sec:process:type:standard}
Participating in the Graphalytics competition executes the benchmark for ecah data set size category.

The competition benchmark evaluates the system performance with six core algorithms, BFS, WCC, PR, CDLP, LCC, SSSP (see \autoref{sec:definition_algorithms}). For each algorithm, the datasets within a given size category are used. A competition benchmark can fall into one of the five target scales: S, M, L, XL, and 2XL+. Each target scale focuses on processing graphs within certain range of data size, and therefore a corresponding time-out duration has been imposed as shown in \autoref{tab:competition-benchmarks}.

\begin{table}[htbp]
	\centering
	\begin{tabular}{|l|r|}
		\hline
		\multicolumn{1}{|c|}{\bf Size} & \multicolumn{1}{c|}{\bf Timeout} \\
		\hline
		S          & 15 minutes    \\
		M          & 30 minutes    \\
		L          & 1 hour        \\
		XL         & 2 hours       \\
		2XL+       & 3 hours       \\
		\hline
	\end{tabular}
	\caption{Benchmarks used in the competition}
	\label{tab:competition-benchmarks}
\end{table}

Each algorithm is executed 3 times.

\section{Benchmark Execution}
\label{sec:process:execution}
\begin{figure}[h]
 	\centering
 	\includegraphics[width=0.9\linewidth]{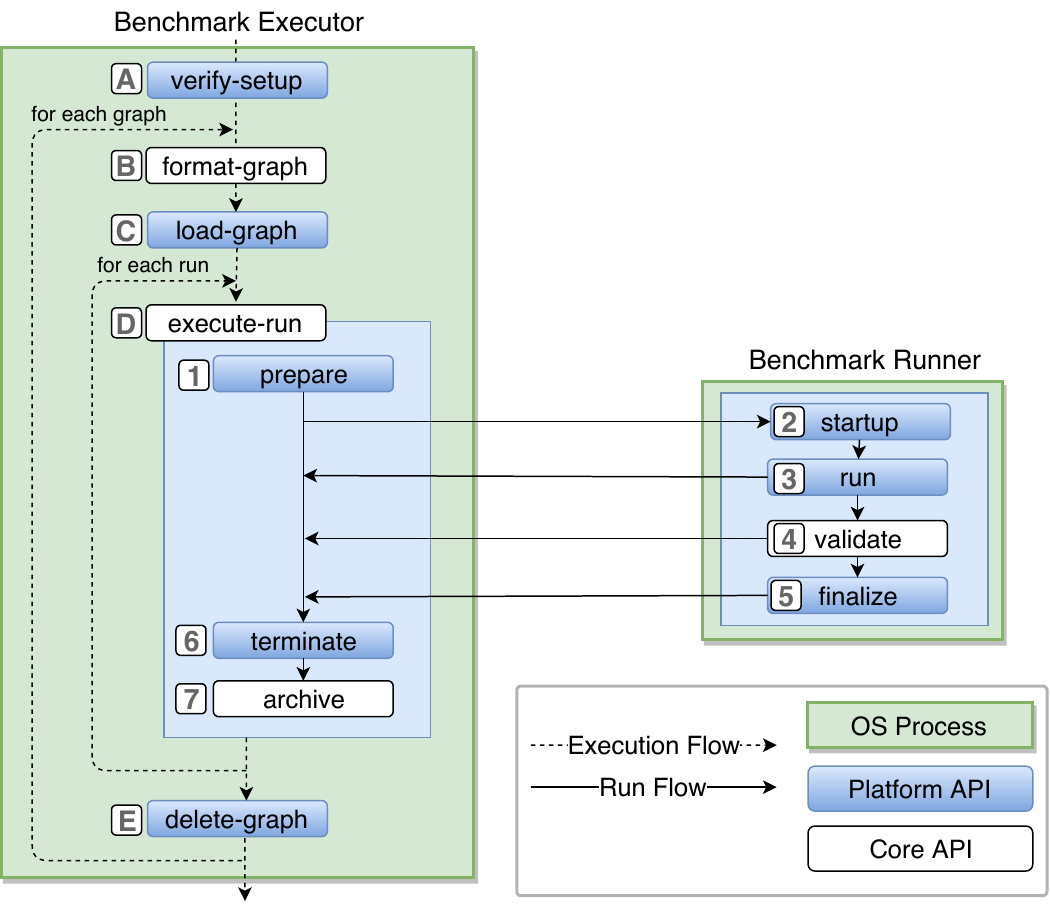}
 	\caption{Benchmark execution in the Graphalytics benchmark suite.}
 	\label{fig:benchmark-process}
\end{figure}

The benchmark execution of Graphalytics benchmark suite is illustrated in \autoref{fig:benchmark-process}. This section explains how the benchmark suite executes a benchmark with regard to its execution flow, run flow, data flow, metric collection, and failure indication.

\subsection{Execution Flow}
\label{sec:process:execution:exe_flow}
After a benchmark is loaded, the benchmark suite analyzes the exact composition of the benchmark. Each benchmark consists of a number of benchmark runs, which will be grouped by the graph dataset used by that benchmark run. For each graph dataset, the input data of that dataset will be loaded only once, and be reused for all corresponding benchmarks runs, before finally being removed.

\begin{enumerate}[label=\textbf{[\Alph*]}]
    \item \textbf{Verify-setup:} The benchmark suite verifies that the platform and the environment are properly set up based on the prerequisites defined in the platform driver.
    
    \item \textbf{Format-graph:} The benchmark suite minimizes the ``input data'' into ``formatted dataset'' (see more in \autoref{sec:process:execution:run_flow}) by removing unused vertex and edge properties.
    
    \item \textbf{Load-graph:} The platform converts the ``formatted data'' into any platform-specific data format and loads a graph dataset into a storage system, which can be either a local file system, a share file system or a distributed file system. This step corresponds to the ``Loading'' step of a graph processing job described in \autoref{sec:def:job:operation}.
    
    \item \textbf{Execute-run:} The platform executes a benchmark run with a specific algorithm and dataset (see more details in \autoref{sec:process:execution:run_flow}). All benchmark runs using the same input dataset can use the prepared graph dataset during the ``load-graph'' step. 
    
    \item \textbf{Delete-graph:} The platform unloads a graph dataset from the storage system, as part of the cleaning up process after all benchmark runs on that graph dataset have been completed.
\end{enumerate}

Note that ``load-graph'' and ``delete-graph'' are platform-specific API, which can be implemented in the platform driver via the ``Platform'' interface, whereas  ``execute-run'' is a uniform step for all platforms. \futureinversion{2.0}{What about format-graph? [Gabor]}

\subsection{Run Flow}
\label{sec:process:execution:run_flow}
The execution of each benchmark run consists of seven steps in total, i.e., ``prepare'', ``startup'', ``run'', ``validate'', ``finalize'', ``terminate'', and  ``archive''. To ensure the stability, the benchmark suite only prepares for the benchmark, and terminates the benchmark run. Each benchmark run is partially executed in an isolated operating-system process, such that a timed-out job can be terminated properly.

\begin{enumerate}[label=\textbf{[\arabic*]}]
    \item \textbf{Prepare:} The platform requests computation resources from the cluster environment and makes the background applications ready.
    
    \item \textbf{Startup:} The platform configures the benchmark run with regard to real-time cluster deployment information, e.g., input directory, output directory and log directory.
    
    \item \textbf{Run:} The platform runs a graph-processing job as defined in the benchmark run. The graph-processing job must complete within the time-out duration, or the benchmark run will fail. This step corresponds to the ``Running'' step of a graph processing job described in \autoref{sec:def:job:operation}.
    
    \item \textbf{Validate:} The benchmark suite validates the platform output with the validation data. The system under test must succeed in this step, or the benchmark run will fail.
    
    \item \textbf{Finalize:} The platform reports the benchmark information and makes the environment ready for the next benchmark run.
    
    \item \textbf{Terminate:} The platform forcibly stops the benchmark job and cleans up the environment, given that the time-out has been reached. 
    
    \item \textbf{Archive:} The benchmark suite archives the benchmark results, gathering information regarding performance metrics and failure indications.
\end{enumerate}

Note that ``prepare'', ``startup'', ``run'', ``finalize'', ``terminate'', and ``archive'' are platform-specific API, which can be implemented in the platform driver via the ``Platform'' interface, whereas ``archive'' is a uniform step for all platforms. \futureinversion{2.0}{The ``archive'' step belongs to the Core API. What about ``validate''? [Gabor]}

\subsection{Data Flow}
\label{sec:process:execution:data_flow}
The graph datasets go through a series of execution steps during the benchmark execution, and in the process of which the format, the representation, and the content of the graph dataset change accordingly.

For each graph, the input datasets and the validation datasets are publicly available benchmark resources.

\begin{itemize} 
    \item \textbf{Input data:} The ``input data'' consists of a vertex file and edge file in EVLP format, as defined in \autoref{sec:data:representation}.
    \item \textbf{Validation data:} The ``validation data'' consists of correct outputs for all six core algorithm, as defined in \autoref{sec:definitions_validation}.
\end{itemize}

The input dataset can be converted into the following format during the benchmark.

\begin{itemize}
    \item \textbf{Formatted data:} The ``input data'' can plausibly contain dozens of vertex and edge properties. During the ``load-graph'' step, the benchmark suite identifies for each algorithms which properties are needed and which are not, and minimizes the ``input data'' into the ``formatted data''. The ``formatted data'' is cached in the storage system for future uses. 
    \item \textbf{Loaded data:} The minimized ``formatted data'' is loaded into a storage system during the ``load-graph'' step, which can either be a local file system, a share file system or a distributed file system.
    \item \textbf{Output data:} The ``output data'' is the output of a graph-processing job being benchmarked during the ``process'' step. The ``output data'' is compared to the ``validation data'' to ensure the correctness of the benchmark execution.
\end{itemize}

\subsection{Failure Indication}
\label{sec:process:execution:failure}
Failures can occur during the benchmark for many reasons. The benchmark suite logs the benchmark execution and classifies the type of failures.

\begin{itemize}
    \item \texttt{DAT}: ``Data failure'' occurs when the ``format-graph`` step fails to generate ``formatted-graph``, or the ``load-graph'' step fails to complete correctly. For example, ``input graph'' can be missing or simply be misplaced, or alternatively the conversion from ``input-graph'' to ``formatted-graph'' could be prematurely interrupted, which leads to data corruption.
    
    \item \texttt{INI}: ``Initialization failure'' occurs when the platform fails to properly make the environment ready for the benchmark during the ``prepare'' or ``startup'' step. For example, the deployment system may fail to allocate the cluster resources needed.
    
    \item \texttt{EXE}: ``Execution failure'' occurs when the execution of the benchmark run fails to complete during the ``run'' step.
    
    \item \texttt{TIM}: ``Time-out failure'' occurs when the pre-defined time-out duration is reached during the ``run'' step.

    \item \texttt{COM}: ``Completion failure'' occurs when output results are incomplete or cannot be found at all. For example, outputs from some compute nodes can be fetched incorrectly.
    
    \item \texttt{VAL}: ``Validation failure'' occurs when the ``output data'' is returned by the system, but fails the validation during the ``validate'' step.
    
    \item \texttt{MET}: ``Metric failure'' occurs when the compulsory performance metrics are missing during the ``archive'' step. For example, the log files containing the information can be non-existing or corrupted.
\end{itemize}

\section{Benchmark Result}

A complete result for the Graphalytics benchmark includes at least the following information: 

\begin{enumerate}
	\item Target scale (T).
	\item Environment specification, including number and type of CPUs, amount of memory, type of network, etc.
	\item Versions of the platform and Graphalytics drivers used in the experiments.
	\item Any non-default configuration options for the platform required to reproduce the system under test.
	\item For every benchmark job:
		\begin{enumerate}
			\item Job specification, i.e., dataset and algorithm.
			\item For every platform run, report the measured processing time, makespan, and whether the run breached the Graphalytics SLA.
			\item (optional) {\tt Granula} archives for each platform run, enabling deep inspection, visualization, modeling, and sharing of performance data.  
		\end{enumerate}
		
\futureinversion{2.0}{
	\item If scalability experiments are performed:
		\begin{enumerate}
			\item Definition of the hardware resources that were scaled up for each scalability experiment.
			\item For every benchmark job corresponding to a scalability experiment, include the resource scale (i.e., 1, 2, 4, 8, or 16).
		\end{enumerate}
}

\futureinversion{2.0}{
	\item 
	If robustness experiments are performed:
		\begin{enumerate}
			\item Summary of the results for every benchmark job.
		\end{enumerate}	
}

\end{enumerate}
Future versions of the benchmark specification will include a Full Disclosure Report template and a process for submitting official Graphalytics results. A sample data format can be found in Appendix~\ref{chap:data-format}.

\chapter{\toolname-based Competitions} \label{chap:competitions}

In this chapter, we describe how to participate in \toolname-based competitions. 
For more details, check out the competition specification document~\cite{CompetitionSpecification}.

\toolname{} defines several {\it official competitions} (\autoref{sec:competitions:overview}), which are open globally to everyone who satisfies the rules for participation. Each competition ranks the benchmark submission based on a competition method defined in \autoref{sec:competitions}.

\section{Official \toolname{} Competitions}
\label{sec:competitions:overview}
Currently, \toolname{} defines two official competitions: (1)~the Global LDBC Competition and (2)~the Global \toolname{} Competition.

\subsection{The Global LDBC Competition} \label{sec:competitions:ldbc}
The Global LDBC Competition is maintained by LDBC, in particular by the Graphalytics Task Force. By the rules of the LDBC charter~\cite{ldbc_byelaws}, the competition method follows the single value-of-merit approach described in \autoref{sec:competitions:single_value}, and focuses on two primary metrics: {\bf ``performance''} and {\bf ``cost-performance''}.

The competition reports the following list of metrics:
\begin{enumerate}
    \item (informative only) Full disclosure of the ``system under test'' (platform + environment).
    \item (informative only) {\it Target scale} of the benchmark.
    \item (informative only) {\it Date} of the benchmark execution.
	\item (flagship value-of-merit) {\it Performance metric}, as summarized from ``EVPS'' of all benchmarked jobs.
	\item (capability value-of-merit) {\it Cost-performance metric}, as summarized from ``PPP'' of all benchmarked jobs.
	\item (informative only) Three {\it performance metrics}, as summarized from $T_l$, $T_m$, and $T_p$ respectively.
\end{enumerate}

\begin{description}
    \item[Maintained by:] LDBC, \url{ldbcouncil.org}.
    \item[Audience:] The LDBC Competition accepts submissions from a global audience.
\end{description} 

\futureinversion{2.0}{Ratio metrics (scalability) and additional energy-related metrics will also be considered.}

\subsection{The Global \toolname{} Competition} \label{sec:competitions:graphalytics}
The Global \toolname{} Competition is maintained by the \toolname team. The competition method follows the tournament-based approach described in \autoref{sec:competitions:tournament}, and focuses on two primary scores: {\bf ``performance''} and {\bf ``cost-performance''}.

The Graphalytics consists of a number of {\it matches}, where each match represents a type of experiment that focuses on a specific performance characteristic that is common across all systems, for example, the EVPS of the BFS algorithm on a Datagen dataset. Each match consists of a set of instances, with the {\bf tournament score} being for each system the sum of {\bf instance scores} accumulated by the platform across all matches in which it participates.  Each {\it instance} is a head-to-head comparison between two systems, for example, comparing the EVPS of any algorithm-dataset for the pair (Giraph, GraphX): the winner receives 1 point, the loser 0 points, and a draw rewards each platform with 0.5 points each.

\begin{enumerate}
    \item (informative only) Full disclosure of the ``system under test'' (platform + environment).
    \item (informative only) {\it Target scale} of the benchmark.
    \item (informative only) {\it Date} of the benchmark execution.
	\item (ranking) {\it Performance score}, by comparing pair-wisely ``EVPS'' of all benchmarked jobs.
	\item (ranking) {\it Cost-performance score}, by comparing pair-wisely  ``PPP'' of all benchmarked jobs.
\end{enumerate}

\begin{description}
    \item[Maintained by:] \toolname, \url{graphalytics.org}.
    \item[Audience:] The Global \toolname{} Competition accepts submissions from a global audience.
\end{description}

\futureinversion{2.0}{Ratio metrics (scalability) and additional energy-related metrics will also be considered.}

\section{Competition Method} \label{sec:competitions}
Different competition methods have been developed to performance comparison in many application domains. Comparing multiple platforms across multiple performance metrics is not trivial. Two major approaches exist for this task: (i) creating a compound metric, typically by weighting the multiple metrics, and comparing multiple platforms using only this single-value-of-merit, and (ii) using a tournament format that allows for multiple participants (platforms) to be compared across multiple criteria. 

The former requires metrics to be easy to compare and compose, that is, to be normalized, to be similarly distributed, to have the same meaning of better (e.g., lower values), to be of importance universally recognized across the field of practice so that weights can be easily ascribed. The latter requires a good tournament format, which does not favor any of the participants, and which does not make participation cumbersome through a large set of rules.

\subsection{Single Value-of-merit Approach} \label{sec:competitions:single_value}
Where metrics (see \autoref{sec:def:metrics}) are collected repeatedly, e.g., each combination of algorithm and dataset, a single value-of-merit can be summarized following the typical processes of benchmarking HPC systems~\cite{DBLP:conf/sc/HoeflerB15}:

\begin{itemize}
	\item For \textbf{Performance metrics,} the \emph{arithmetic mean} across all data.
	\item For \textbf{Throughput metrics,} because they are rate metrics, in two consecutive steps:
		\begin{enumerate}
			\item let $a$ be the \emph{arithmetic mean} of the performance metric (e.g., processing time) and $w$ be the (constant, total) workload (e.g., count of edges plus vertices),
			\item report the \emph{ratio} between $w$ and $a$ as the throughput metric.
		\end{enumerate}
 		In other words, instead of averaging the rate per sample, that is, $\textit{EVPS}_i$ for sample $i$, \toolname{} first averages the performance metric and then reports the rate. \futureinversion{2.0}{Maybe change ``rate'' to ``ratio'' in this sentence [Gabor]}
	\item For \textbf{Cost metrics,} the \emph{harmonic mean} across all data. This is because the denominator (e.g., EVPS for PPP) gives meaning to the ratio (TCO is constant across experiments with the same System Under Test), which indicates that the arithmetic mean would be misleading~\cite[S.3.1.1]{DBLP:conf/sc/HoeflerB15}.
	\item For \textbf{Ratio metrics} such as Speedup, the \emph{geometric mean} across all data.
\end{itemize}

\subsection{Tournament-based Approach} \label{sec:competitions:tournament}
In a tournament-based approach, the system performance is ranked by means of competitive tournaments~\cite{Thurstone1927}.  Generally, a Round-Robin pair-wise tournament~\cite{David1960} (from hereon, {\it tournament}) of $p$ participants involves a balanced set of (pair-wise) comparisons between the results of each pair of participants; if there are $c$ criteria to compare the participants, there will be $\frac{1}{2} \times c \times p (p - 1)$ pair-wise comparisons. In a pair-wise comparison, a pre-defined amount of points (often, 1 or 3) is given to the better ({\it winner}) participant from the pair. It is also common to give zero points to the worse ({\it loser}) participant from the pair, and to split the points between participants with equal performance. Similar tournaments have been used for decades in chess competitions, in professional sports leagues such as (European and American) football, etc.

We do not consider here other pair-wise tournaments, such as replicated tournaments~\cite{David1960} and unbalanced comparisons~\cite{david1987ranking}, which have been used especially in settings where comparisons are made by human referees and are typically discretized on 5-point Likert scales, and thus are quantitatively less accurate than the \toolname{} measurements.

\chapter{Implementation Instructions}
\label{chap:instructions}
Graphalytics provides a set of benchmark software and resources which are open-source and publicly available. This chapter explains how to work with Graphalytics software and enumerates the available benchmark resources which are necessary for the benchmark.

\section{Graphalytics Software and Documentation}\label{sec:instructions:core}
The Graphalytics team develops and maintains the core Graphalytics software, which facilitates the benchmark process and is extendable for benchmarking and analyzing the performance of various graph processing platforms.

\paragraph{Graphalytics Core}
The Graphalytics Core contains the core implementation for the Graphalytics benchmark, provides a programmable interface for platform drivers.

\quad Link: \url{https://github.com/ldbc/ldbc_graphalytics}

\paragraph{Graphalytics Specification}
The source code for generating this specification.

\quad Link: \url{https://github.com/ldbc/ldbc_graphalytics_docs}

\paragraph{Reference Implementations}

We provide two reference implementations.

\begin{itemize}
	\item SuiteSparse:GraphBLAS~\cite{DBLP:journals/toms/Davis19}: \url{https://github.com/ldbc/ldbc_graphalytics_platforms_graphblas}
	\item Umbra~\cite{DBLP:conf/cidr/NeumannF20}: \url{https://github.com/ldbc/ldbc_graphalytics_platforms_umbra}
\end{itemize}

\printbibliography

\appendix
\chapter{Pseudo-code for Algorithms}
\label{chap:algorithms}
This chapter contains pseudo-code for the algorithms described in \autoref{sec:definition_algorithms}. In the following sections, a graph $G$ consists of a set of vertices $V$ and a set of edges $E$. For undirected graphs, each edge is bidirectional, so if $(u, v) \in E$ then $(v, u) \in E$. Each vertex has a set of outgoing neighbors $N_\mathrm{out}(v) = \{u \in V | (v, u) \in E\}$ and a set of incoming neighbors $N_\mathrm{in}(v) = \{u \in V | (u, v) \in E\}$.

\section{Breadth-First Search (BFS)}

\begin{algorithm}[h!]
\begin{algorithmic}[1]
\Statex \textbf{input:} graph $G=(V,E)$, vertex $\textit{root}$
\Statex \textbf{output:} array $\textit{depth}$ storing vertex depths
\ForAll{$v \in V$}
  \State $\textit{depth}[v] \gets \infty$
\EndFor
\State $Q$ $\gets$  \textsc{create\_queue()}
\State \Call{$Q$.push}{$\textit{root}$}
\State $\textit{depth}[\textit{root}] \gets 0$
\While{\Call{$Q$.size}{} > 0}
  \State $v \gets $ \Call{$Q$.pop\_front}{ }
  \ForAll{$u \in N_\mathrm{out}(v)$}
    \If{$\textit{depth}[u] = \infty$}
      \State $\textit{depth}[u] \gets \textit{depth}[v] + 1$
      \State \Call{$Q$.push\_back}{$u$}
    \EndIf
  \EndFor
\EndWhile
\end{algorithmic}
\end{algorithm}

\section{PageRank (PR)}

\begin{algorithm}[h!]
\begin{algorithmic}[1]
\Statex \textbf{input:} graph $G=(V,E)$, integer $\textit{max\_iterations}$
\Statex \textbf{output:} array $\textit{rank}$ storing PageRank values
\ForAll{$v \in V$}
  \State $\textit{rank}[v] \gets \frac{1}{|V|}$
\EndFor
\For{$i=1,\ldots,\textit{max\_iterations}$}
\State $\textit{sink\_sum} \gets 0$
\ForAll{$w \in V$}
  \If{$|N_\mathrm{out}(w)| = 0$}
    \State $\textit{sink\_sum} \gets \textit{sink\_sum} + \textit{rank}[w]$
  \EndIf
\EndFor
\ForAll{$v \in V$}
  \State $\textit{new\_rank}[v] \gets \frac{1-d}{|V|} + d \cdot \sum_{u \in N_\mathrm{in}(v)} \frac{\textit{rank}[u]}{|N_\mathrm{out}(u)|} + \frac{d}{|V|} \cdot \textit{sink\_sum} $
\EndFor
\State $\textit{rank} \gets \textit{new\_rank}$
\EndFor
\end{algorithmic}
\end{algorithm}

\clearpage

\section{Weakly Connected Components (WCC)}

\begin{algorithm}[h!]
\begin{algorithmic}[1]
\Statex \textbf{input:} graph $G=(V,E)$
\Statex \textbf{output:} array $\textit{comp}$ storing component labels
\ForAll{$v \in V$}
  \State $\textit{comp}[v] \gets v$
\EndFor
\Repeat
\State $\textit{converged} \gets \text{true}$
\ForAll{$v \in V$}
  \ForAll{$u \in N_\mathrm{in}(v) \cup N_\mathrm{out}(v)$}
    \If{$\textit{comp}[v] > \textit{comp}[u]$}
      \State $\textit{comp}[v] \gets \textit{comp}[u]$
      \State $\textit{converged} \gets \text{false}$
     \EndIf
  \EndFor
\EndFor
\Until{$\textit{converged}$}
\end{algorithmic}
\end{algorithm}

\section{Local Clustering Coefficient (LCC)}

\begin{algorithm}[h!]
\begin{algorithmic}[1]
\Statex \textbf{input:} graph $G=(V,E)$
\Statex \textbf{output:} array $\textit{lcc}$ storing LCC values
\ForAll{$v \in V$}
   \State $d \gets |N_\mathrm{in}(v) \cup N_\mathrm{out}(v)|$
  \If {$d \geq 2$}
  \State $t \gets 0$
  \ForAll{$u \in N_\mathrm{in}(v) \cup N_\mathrm{out}(v)$}
    \ForAll{$w \in N_\mathrm{in}(v) \cup N_\mathrm{out}(v)$}
     \If{$(u, w) \in E$}
      \Comment{Check if edge $(u, w)$ exists}
      \State $t  \gets t + 1$
      \Comment{Found triangle $v-u-w$}
     \EndIf
    \EndFor
  \EndFor
    \State $\textit{lcc}[v] \gets \frac{t}{d(d-1)} $
  \Else
    \State $\textit{lcc}[v] \gets 0$
    \Comment{No triangles possible}
  \EndIf
\EndFor
\end{algorithmic}
\end{algorithm}

\clearpage

\section{Community Detection using Label Propagation (CDLP)}

\begin{algorithm}[h!]
\begin{algorithmic}[1]
\Statex \textbf{input:} graph $G=(V,E)$, integer $\textit{max\_iterations}$
\Statex \textbf{output:} array $\textit{labels}$ storing vertex communities
\ForAll{$v \in V$}
  \State $\textit{labels}[v] \gets v$
\EndFor
\For{$i=1, \ldots, \textit{max\_iterations}$}
 \ForAll{$v \in V$}
  \State $C$ $\gets$ \textsc{create\_histogram()}

  \ForAll{$u \in N_\mathrm{in}(v)$}
    \State \Call{$C$.add}{$\textit{labels}[u]$}
  \EndFor
  \ForAll{$u \in N_\mathrm{out}(v)$}
    \State \Call{$C$.add}{$\textit{labels}[u]$}
  \EndFor
  \State $\textit{freq} \gets $ \Call{$C$.get\_maximum\_frequency}{ }
  \Comment{Find maximum frequency of labels}
  \State $\textit{candidates} \gets$ \Call{$C$.get\_labels\_for\_frequency}{$\textit{freq}$}
  \Comment{Find labels with max.\ frequency}
  \State $\textit{new\_labels}[v] \gets$ \Call{min}{$\textit{candidates}$}
  \Comment{Select smallest label}
 \EndFor
 \State $\textit{labels} \gets \textit{new\_labels}$
\EndFor
\end{algorithmic}
\end{algorithm}

\section{Single-Source Shortest Paths (SSSP)}

\begin{algorithm}[h!]
\begin{algorithmic}[1]
\Statex \textbf{input:} graph $G=(V,E)$, vertex $\textit{root}$, edge weights $\textit{weight}$
\Statex \textbf{output:} array $\textit{dist}$ storing distances
\ForAll{$v \in V$}
  \State $\textit{dist}[v] \gets \infty$
\EndFor

\State $H$ $\gets \Call{create\_heap}{ }$
\State \Call{$H$.insert}{root, 0}
\State $\textit{dist}[\textit{root}] \gets 0$
\While{$\Call{$H$.size}{} > 0$}
  \State $v \gets$ \Call{$H$.delete\_minimum}{ }
  \Comment{Find vertex $v$ in $H$ such that $\textit{dist}[v]$ is minimal}
  \ForAll{$w \in N_\mathrm{out}(v)$}
    \If{$\textit{dist}[w] > \textit{dist}[v] + \textit{weight}[v,w]$}
      \State $\textit{dist}[w] \gets \textit{dist}[v] + \textit{weight}[v,w]$
      \State \Call{$H$.insert}{$w$, $\textit{dist}[w]$}
    \EndIf
  \EndFor
\EndWhile
\end{algorithmic}
\end{algorithm}

\chapter{Data format for Benchmark Results}
\label{chap:data-format}

This appendix shows an example of Graphalytics benchmark results of the reference implementation. Graphalytics benchmark defines a specific data format for the benchmark results. The result is formatted in JSON, and consists of three main components: system under test, benchmark configuration, and experimental results. \autoref{fig:result-format:overview} depicts the top-level structure of the result format.

\begin{lstlisting}[language={json},caption={Result Format: Overview},label={fig:result-format:overview}]
{
  "id": "b2940223",
  "system": {...},
  "configuration": {...},
  "result": {...}
}
\end{lstlisting}

For the system under test, Graphalytics reports the detailed descriptions of the graph analytic platform, the cluster environment, and the benchmark tool.

\begin{lstlisting}[language={json},caption={Result Format: System Under Test},label={fig:result-format:system}]
"system": {
  "platform": {
    "name": "reference",
    "acronym": "ref",
    "version": "1.4.0",
    "link": "https://github.com/ldbc/ldbc_graphalytics_platforms_reference"
  },
  "environment": {
    "name": "DAS Supercomputer",
    "acronym": "das5",
    "version": "5",
    "link": "http://www.cs.vu.nl/das5/",
    "machines": [
      {
        "quantity": 20,
        "operating-system": "Centos",
        "cpu": { 
          "name": "XEON", 
          "cores": "16"
        },
        "memory": {
          "name": "?",
          "size": "40GB"
        },
        "network": {
          "name": "Infiniband",
          "throughput": "10GB/s"
        },
        "storage": {
          "name": "SSD",
          "volume": "20TB"
        },
        "accel": {}
      }
    ]
  },
  "benchmark": {
    "graphalytics-core": {
      "name": "graphalytics-core",
      "version": "1.1.0",
      "link": "https://github.com/ldbc/ldbc_graphalytics"
    },
    "graphalytics-platforms-reference": {
      "name": "graphalytics-platforms-reference",
      "version": "2.1.0",
      "link": "https://github.com/ldbc/ldbc_graphalytics_platforms_reference"
    }
  }
}
\end{lstlisting}

For the benchmark configuration, the target scale and the computation resource usage is reported. For each resource type, the baseline resource usage, and the scalablity of that resource type is reported.

\begin{lstlisting}[language={json},caption={Result Format: Benchmark Configuration},label={fig:result-format:conf}]
"configuration": {
  "target-scale": "L",
  "resources": {
    "cpu-instance": { "name": "cpu-instance", "baseline": 1,  "scalability": true  },
    "cpu-core":     { "name": "cpu-core",     "baseline": 32, "scalability": false },
    "memory":       { "name": "memory",       "baseline": 64, "scalability": true  },
    "network":      { "name": "network",      "baseline": 10, "scalability": false }
  }
}
\end{lstlisting}

\begin{lstlisting}[language={json},caption={Result Format: Experiment Result},label={fig:result-format:result}]
"result": {
  "experiments": {
    "e34252": {
      "id": "e34252",
      "type": "baseline-alg-bfs",
      "jobs": [ "j34252", "j75352", "j23552" ]
    },
    "e75352": {
      "id": "e75352",
      "type": "baseline-alg-pr",
      "jobs": [ "j34252", "j75352", "j23552" ]
    },
    "e23552": {
      "id": "e23552",
      "type": "baseline-alg-cdlp",
      "jobs": [ "j34252", "j75352", "j23552" ]
    }
  },
  "jobs": {
    "j34252": {
      "id": "j34252",
      "algorithm": "bfs",
      "dataset": "D100",
      "scale": 1,
      "repetition": 3,
      "runs": [ "r649352", "r124252", "r124252" ]
    },
    "j75352": {
      "id": "j75352",
      "algorithm": "pr",
      "dataset": "D1000",
      "scale": 1,
      "repetition": 3,
      "runs": [ "r649352", "r124252", "r124252" ]
    },
    "j23552": {
      "id": "j23552",
      "algorithm": "cdlp",
      "dataset": "G25",
      "scale": 1,
      "repetition": 3,
      "runs": [ "r649352", "r124252", "r124252" ]
    }
  },
  "runs": {
    "r649352": {
      "id": "r649352",
      "timestamp": 1463310828849,
      "success": true,
      "makespan": 23423422,
      "processing-time": 2234
    },
    "r124252": {
      "id": "r124252",
      "timestamp": 1463310324849,
      "success": true,
      "makespan": 2343422,
      "processing-time": 234
    },
    "r643252": {
      "id": "r124252",
      "timestamp": 1463310324849,
      "success": true,
      "makespan": 2343422,
      "processing-time": 234
    }
  }
}
\end{lstlisting}

For the experimental results, the set of experiments, the underlying jobs, and the corresponding runs are reported.

\chapter{Related work}
\label{chap:related-work}

\begin{table*}[!ht]
	\centering
	\caption[Overview of related work]{Overview of related work. {\small (Acronyms: 
			\emph{Reference type}: {\bf S}, study; {\bf B}, benchmark. \emph{Target system, structure}: {\bf D}, distributed system; {\bf P}, parallel system; {\bf MC},  single-node multi-core system; {\bf GPU}, using GPUs. 
			\emph{Input}: {\bf 0}, no parameters; {\bf S}, parameters define scale; {\bf E}, parameters define edge properties; {\bf +}, parameters define other graph properties, e.g., clustering coefficient.
			\emph{Datasets/Algorithms}: {\bf Rnd}, reason for selection not explained; {\bf Exp}, selection guided by expertise; {\bf 1-stage}, data-driven selection; {\bf 2-stage}, 2-stage data- and expertise-driven process.
			\emph{Scalability tests}: {\bf W}, weak; {\bf S}, strong; {\bf V}, vertical; {\bf H}, horizontal.)}}
	\label{tab:SummaryOfRelatedWorkAcronyms}
	\resizebox{\textwidth}{!}{
		\begin{tabular}[!tb]{|l|l|l|l|l|l|l|l|l|l|l|}
\hline
\multicolumn{2}{|c|}{Reference (chronological order)} & \multicolumn{2}{c|}{Target System (R1)} & \multicolumn{4}{c|}{Design (R2)} & \multicolumn{2}{c|}{Tests (R3)} & \multicolumn{1}{c|}{(R4)}\\ 
\hline
 & Name [Publication] & \multicolumn{1}{c|}{Structure} & \multicolumn{1}{c|}{Programming} & \multicolumn{1}{c|}{Input} & \multicolumn{1}{c|}{Datasets} & \multicolumn{1}{c|}{Algo.} & \multicolumn{1}{c|}{Scalable?} & \multicolumn{1}{c|}{Scalability} & \multicolumn{1}{c|}{Robustness} & \multicolumn{1}{c|}{Renewal}\\ 
\hline
B & \pbox{5cm}{CloudSuite~\cite{DBLP:conf/asplos/FerdmanAKVAJKPAF12}, \\only graph elements} & \multicolumn{1}{c|}{D/MC} & \multicolumn{1}{c|}{PowerGraph} & \multicolumn{1}{c|}{S} & \multicolumn{1}{c|}{Rnd} & \multicolumn{1}{c|}{Exp} & \multicolumn{1}{c|}{---} & \multicolumn{1}{c|}{No} & \multicolumn{1}{c|}{No} & \multicolumn{1}{c|}{No} \\ 
\hline
S & Montresor et al.~\cite{DBLP:conf/bigdataconf/ElserM13} & \multicolumn{1}{c|}{D/MC} & \multicolumn{1}{c|}{3 classes} & \multicolumn{1}{c|}{0} & \multicolumn{1}{c|}{Rnd} & \multicolumn{1}{c|}{Exp} & \multicolumn{1}{c|}{---} & \multicolumn{1}{c|}{No} & \multicolumn{1}{c|}{No} & \multicolumn{1}{c|}{No} \\ 
\hline
B & HPC-SGAB~\cite{DBLP:conf/hipc/BaderM05} & \multicolumn{1}{c|}{P} & \multicolumn{1}{c|}{---} & \multicolumn{1}{c|}{S} & \multicolumn{1}{c|}{Exp} & \multicolumn{1}{c|}{Exp} & \multicolumn{1}{c|}{---} & \multicolumn{1}{c|}{No} & \multicolumn{1}{c|}{No} & \multicolumn{1}{c|}{No} \\ 
\hline
B & Graph500 & \multicolumn{1}{c|}{P/MC/GPU} & \multicolumn{1}{c|}{---} & \multicolumn{1}{c|}{S} & \multicolumn{1}{c|}{Exp} & \multicolumn{1}{c|}{Exp} & \multicolumn{1}{c|}{---} & \multicolumn{1}{c|}{No} & \multicolumn{1}{c|}{No} & \multicolumn{1}{c|}{No} \\ 
\hline
B & GreenGraph500 & \multicolumn{1}{c|}{P/MC/GPU} & \multicolumn{1}{c|}{---} & \multicolumn{1}{c|}{S} & \multicolumn{1}{c|}{Exp} & \multicolumn{1}{c|}{Exp} & \multicolumn{1}{c|}{---} & \multicolumn{1}{c|}{No} & \multicolumn{1}{c|}{No} & \multicolumn{1}{c|}{No} \\ 
\hline
B & WGB~\cite{DBLP:conf/wbdb/AmmarO13} & \multicolumn{1}{c|}{D} & \multicolumn{1}{c|}{---} & \multicolumn{1}{c|}{SE+} & \multicolumn{1}{c|}{Exp} & \multicolumn{1}{c|}{Exp} & \multicolumn{1}{c|}{1B Edges} & \multicolumn{1}{c|}{No} & \multicolumn{1}{c|}{No} & \multicolumn{1}{c|}{No} \\ 
\hline
\hline
S & {\bf Own prior work}~\cite{DBLP:conf/ipps/GuoBVIMW14,DBLP:conf/ccgrid/GuoVIE15,DBLP:conf/sigmod/CapotaHIPEB14} & \multicolumn{1}{c|}{{\bf D/MC/GPU}} & \multicolumn{1}{c|}{{\bf 10 classes}} & \multicolumn{1}{c|}{{\bf S}} & \multicolumn{1}{c|}{{\bf Exp}} & \multicolumn{1}{c|}{{\bf 1-stage}} & \multicolumn{1}{c|}{{\bf 1B Edges}} & \multicolumn{1}{c|}{{\bf W/S/V/H}} & \multicolumn{1}{c|}{{\bf No}} & \multicolumn{1}{c|}{{\bf No}}\\ 
\hline
\hline
S & \"Ozsu et al.~\cite{DBLP:journals/pvldb/HanDAOWJ14} & \multicolumn{1}{c|}{D} & \multicolumn{1}{c|}{Pregel} & \multicolumn{1}{c|}{0} & \multicolumn{1}{c|}{Exp,Rnd} & \multicolumn{1}{c|}{Exp} & \multicolumn{1}{c|}{---} & \multicolumn{1}{c|}{W/S/V/H} & \multicolumn{1}{c|}{No} & \multicolumn{1}{c|}{No} \\ 
\hline
B & \pbox{5cm}{BigDataBench~\cite{DBLP:conf/wbdb/MingLGHYWZ13,DBLP:conf/hpca/WangZLZYHGJSZZLZLQ14}, \\only graph elements} & \multicolumn{1}{c|}{D/MC} & \multicolumn{1}{c|}{Hadoop} & \multicolumn{1}{c|}{S} & \multicolumn{1}{c|}{Rnd} & \multicolumn{1}{c|}{Rnd} & \multicolumn{1}{c|}{---} & \multicolumn{1}{c|}{S} & \multicolumn{1}{c|}{No} & \multicolumn{1}{c|}{No} \\ 
\hline
S & Satish et al.~\cite{DBLP:conf/sigmod/SatishSPSPHSYD14} & \multicolumn{1}{c|}{D/MC} & \multicolumn{1}{c|}{6 classes} & \multicolumn{1}{c|}{S} & \multicolumn{1}{c|}{Exp,Rnd} & \multicolumn{1}{c|}{Exp} & \multicolumn{1}{c|}{---} & \multicolumn{1}{c|}{W} & \multicolumn{1}{c|}{No} & \multicolumn{1}{c|}{No}\\ 
\hline
S & Lu et al.~\cite{DBLP:journals/pvldb/LuCYW14} & \multicolumn{1}{c|}{D} & \multicolumn{1}{c|}{4 classes} & \multicolumn{1}{c|}{S} & \multicolumn{1}{c|}{Exp,Rnd} & \multicolumn{1}{c|}{Exp} & \multicolumn{1}{c|}{---} & \multicolumn{1}{c|}{S} & \multicolumn{1}{c|}{No} & \multicolumn{1}{c|}{No} \\ 
\hline
B & GraphBIG~\cite{DBLP:conf/sc/NaiXTKL15} & \multicolumn{1}{c|}{P/MC/GPU} & \multicolumn{1}{c|}{System G} & \multicolumn{1}{c|}{S} & \multicolumn{1}{c|}{Exp} & \multicolumn{1}{c|}{Exp} & \multicolumn{1}{c|}{---} & \multicolumn{1}{c|}{No} & \multicolumn{1}{c|}{No} & \multicolumn{1}{c|}{No} \\ 
\hline
S & Cherkasova et al.~\cite{DBLP:conf/wosp/EisenmanCMCFK16} & \multicolumn{1}{c|}{MC} & \multicolumn{1}{c|}{Galois} & \multicolumn{1}{c|}{0} & \multicolumn{1}{c|}{Rnd} & \multicolumn{1}{c|}{Exp} & \multicolumn{1}{c|}{---} & \multicolumn{1}{c|}{No} & \multicolumn{1}{c|}{No} & \multicolumn{1}{c|}{No} \\ 
\hline
\hline
{\bf B} & {\bf LDBC Graphalytics} (this work) & \multicolumn{1}{c|}{{\bf D/MC/GPU}} & \multicolumn{1}{c|}{{\bf 10+ classes}} & \multicolumn{1}{c|}{{\bf SE+}} & \multicolumn{1}{c|}{{\bf 2-stage}} & \multicolumn{1}{c|}{{\bf 2-stage}} & \multicolumn{1}{c|}{{\bf Process}} & \multicolumn{1}{c|}{{\bf W/S/V/H}} & \multicolumn{1}{c|}{{\bf Yes}} & \multicolumn{1}{c|}{{\bf Yes}} \\ 
\hline
\end{tabular}

	}
	
\end{table*}

\autoref{tab:SummaryOfRelatedWorkAcronyms}, which is reproduced from~\cite{DBLP:journals/pvldb/IosupHNHPMCCSATXNB16}, summarizes and compares Graphalytics with previous studies and benchmarks for graph analysis systems. R1--R5 are the requirements formulated in \autoref{sec:requirements}. 
As the table indicates, there is no alternative to Graphalytics in covering requirements R1--R4. We also could not find evidence of requirement R5 being covered by other systems than LDBC.
While there have been a few related \emph{benchmark proposals} (marked ``B''), these either do not {\em focus} on graph analysis, or are much narrower in scope (e.g., only BFS for Graph500).
There have been comparable \emph{studies} (marked ``S'') but these have not attempted to define---let alone maintain---a benchmark, its specification, software, testing tools and practices, or results.
Graphalytics is not only industry-backed but also has industrial strength, through its detailed execution process, its metrics that characterize robustness in addition to scalability, and a renewal process that promises longevity.
Graphalytics is being proposed to SPEC as well, and BigBench~\cite{DBLP:conf/sigmod/GhazalRHRPCJ13,DBLP:conf/sigmod/RablFDJG15} explicitly refers to Graphalytics as its option for future benchmarking of graph analysis platforms.

Previous studies typically tested the open-source platforms Giraph~\cite{DBLP:books/sp/SOAK2016}, GraphX~\cite{DBLP:conf/sigmod/XinGFS13}, and 
PowerGraph~\cite{DBLP:conf/osdi/GonzalezLGBG12}, but our contribution here is that vendors (Oracle, Intel, IBM) in our evaluation have themselves tuned and tested their implementations for PGX~\cite{DBLP:conf/sc/HongDMLVC15}, GraphMat~\cite{DBLP:journals/pvldb/SundaramSPDAV0D15} and OpenG~\cite{DBLP:conf/sc/NaiXTKL15}. We are aware that the database community has started to realize that with some enhancements, RDBMS technology could also be a contender in this area~\cite{DBLP:conf/cidr/FanRP15,DBLP:journals/pvldb/JindalR0MDS14}, and we hope that such systems will soon get tested with Graphalytics. 

Graphalytics complements the many existing efforts focusing on graph databases, such as 
LinkBench~\cite{DBLP:conf/sigmod/ArmstrongPBC13}, 
XGDBench~\cite{DBLP:journals/ase/DayarathnaS14}, and 
LDBC SNB~\cite{DBLP:conf/sigmod/ErlingALCGPPB15,DBLP:conf/grades/SzarnyasPAMPKEB18};  
efforts focusing on RDF graph processing, such as 
LUBM~\cite{DBLP:journals/ws/GuoPH05},
the Berlin {SPARQL} Benchmark~\cite{DBLP:journals/ijswis/BizerS09},
SP\textsuperscript{2}Bench~\cite{DBLP:conf/icde/SchmidtHLP09},
and WatDiv~\cite{DBLP:conf/semweb/AlucHOD14} (targeting also graph databases);
and community efforts such as the TPC benchmarks.
Whereas all these prior efforts are interactive database query benchmarks, Graphalytics focuses on algorithmic graph analysis and on different platforms which are not necessarily database systems, whose distributed and highly parallel aspects lead to different design trade-offs.

The GAP Benchmark Suite~\cite{DBLP:journals/corr/BeamerAP15} targets six graph kernels: BFS, SSSP, PR, WCC, triangle count, and betweenness centrality. The first four are present in Graphalytics, while the \emph{triangle count} kernel shares many challenges with the LCC algorithm. (\emph{Betweenness centrality} was also considered for Graphalytics but it necessitates using approximation methods for large graphs, therefore automated validation of results is not possible.)

\chapter{Example Graphs for Validation}
\label{chap:validation_examples}

In this chapter, we provide test graphs containing the expected results for
BFS (\autoref{fig:bfs_example}),
CDLP (\autoref{fig:cdlp_example}),
LCC (\autoref{fig:lcc_example}),
SSSP (\autoref{fig:sssp_example}),
WCC (\autoref{fig:wcc_example}),
and a common example for all six algorithms (\autoref{fig:common_example}).
The PageRank test graphs (\texttt{test-pr-directed} and \texttt{test-pr-undirected}) are provided in the data sets but not displayed here due to their relatively large sizes (50~nodes).

\newcommand{\examplescale}{0.48}

\begin{figure}[h]
	\centering
	\begin{subfigure}{0.496\textwidth}
		\centering
		\includegraphics[scale=\examplescale]{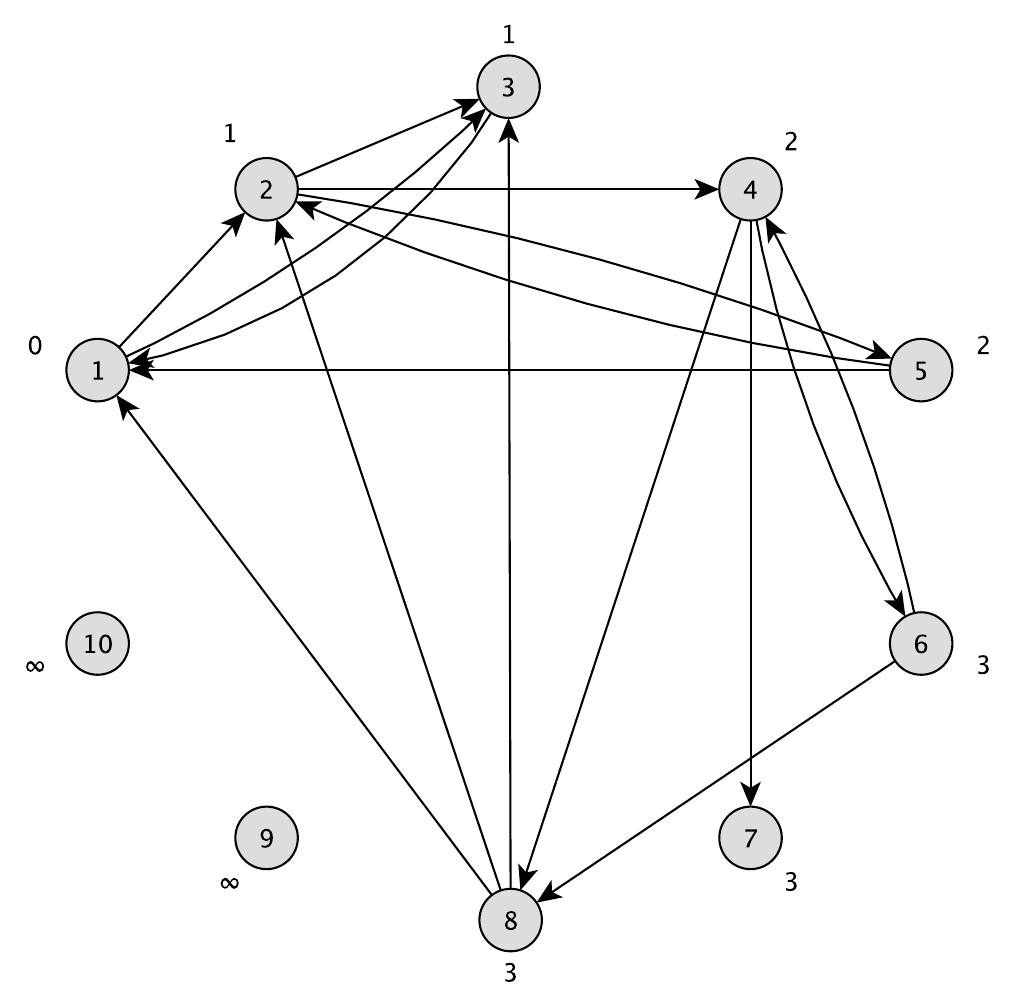}
		\caption{Directed (\texttt{text-bfs-directed})}
	\end{subfigure}
	\begin{subfigure}{0.496\textwidth}
		\centering
		\includegraphics[scale=\examplescale]{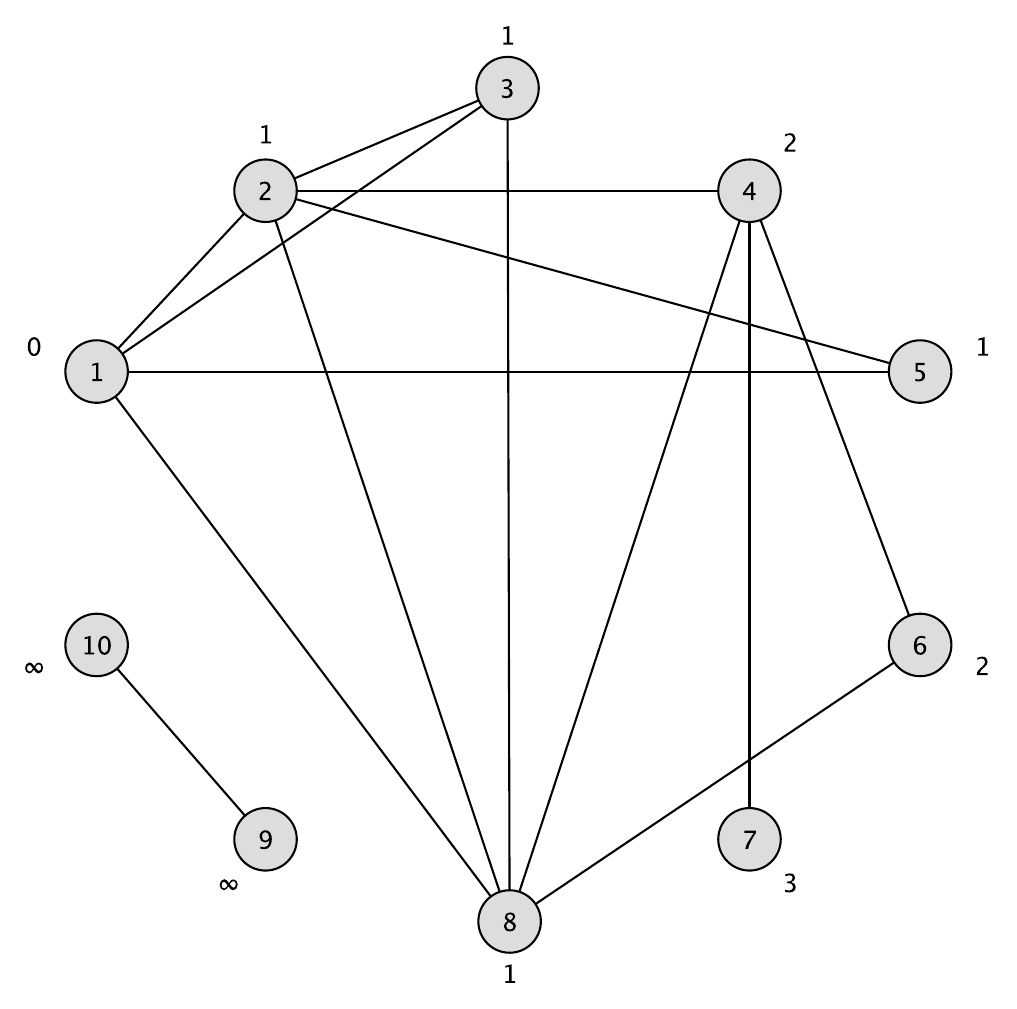}
		\caption{Undirected (\texttt{text-bfs-undirected})}
	\end{subfigure}
	\caption{Test graphs for BFS. Source vertex: 1.}
	\label{fig:bfs_example}
\end{figure}

\begin{figure}[h]
	\centering
	\begin{subfigure}{0.496\textwidth}
		\centering
		\includegraphics[scale=\examplescale]{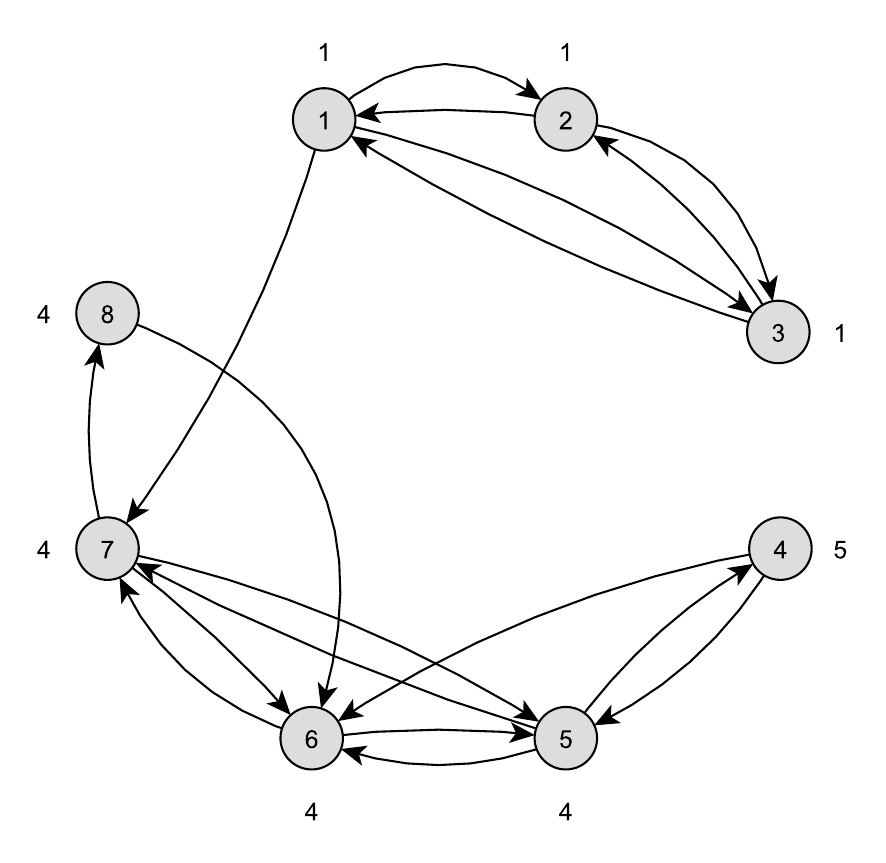}
		\caption{Directed (\texttt{text-cdlp-directed})}
	\end{subfigure}
	\begin{subfigure}{0.496\textwidth}
		\centering
		\includegraphics[scale=\examplescale]{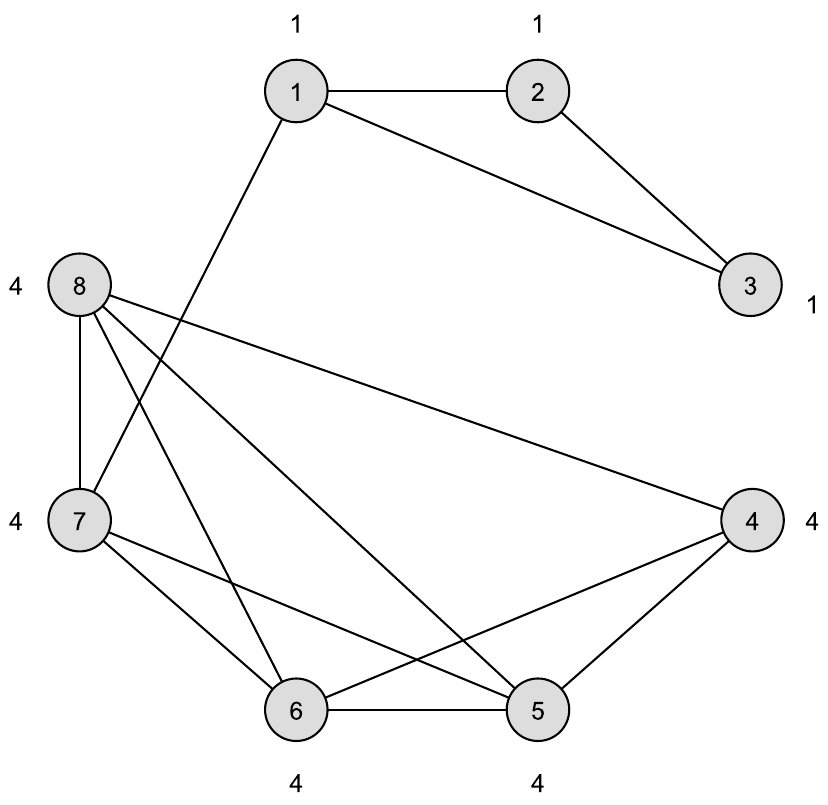}
		\caption{Undirected (\texttt{text-cdlp-undirected})}
	\end{subfigure}
	\caption{Test graphs for CDLP. Maximum number of iterations: 5.}
	\label{fig:cdlp_example}
\end{figure}

\begin{figure}[h]
	\centering
	\begin{subfigure}{0.496\textwidth}
		\centering
		\includegraphics[scale=\examplescale]{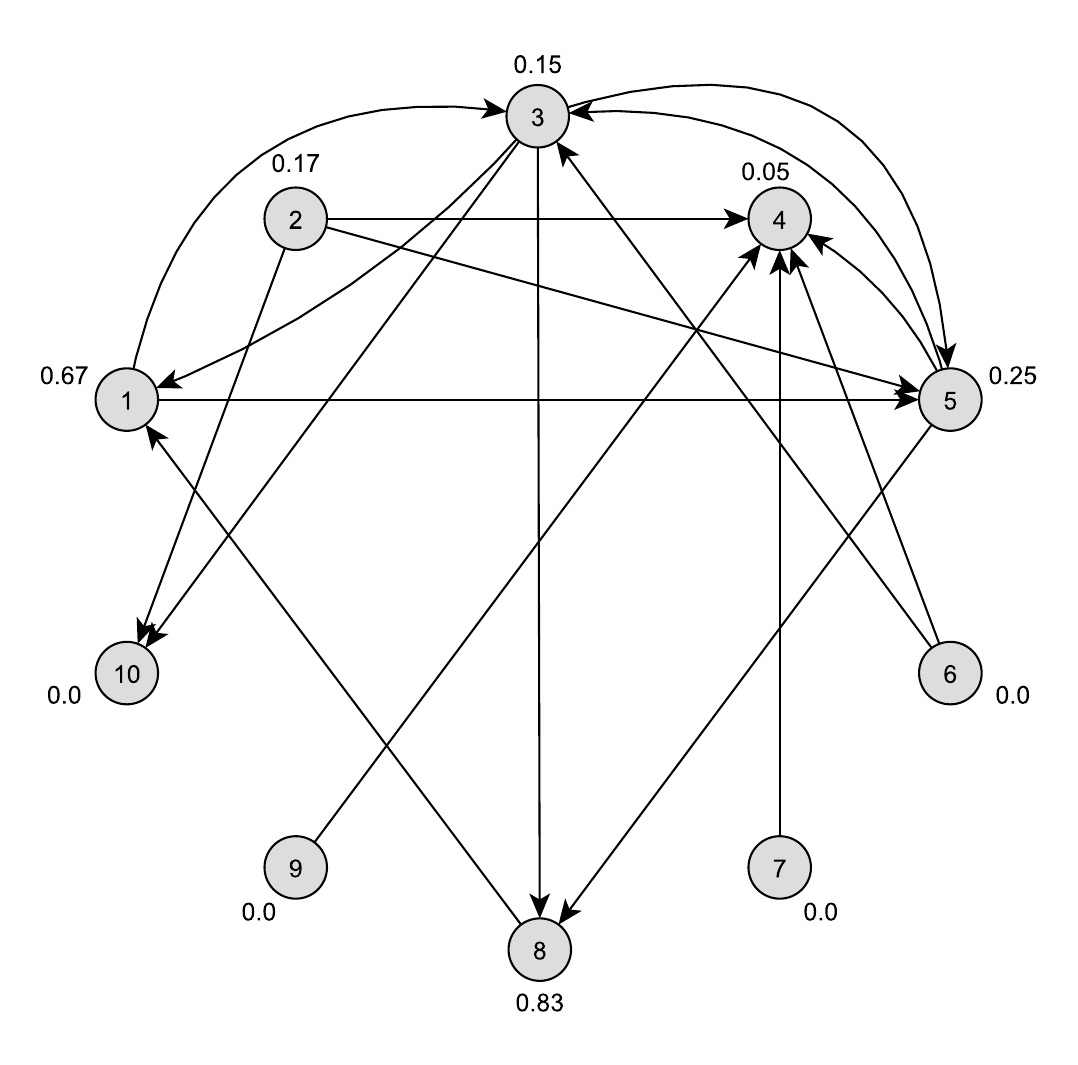}
		\caption{Directed (\texttt{text-lcc-directed})}
	\end{subfigure}
	\begin{subfigure}{0.496\textwidth}
		\centering
		\includegraphics[scale=\examplescale]{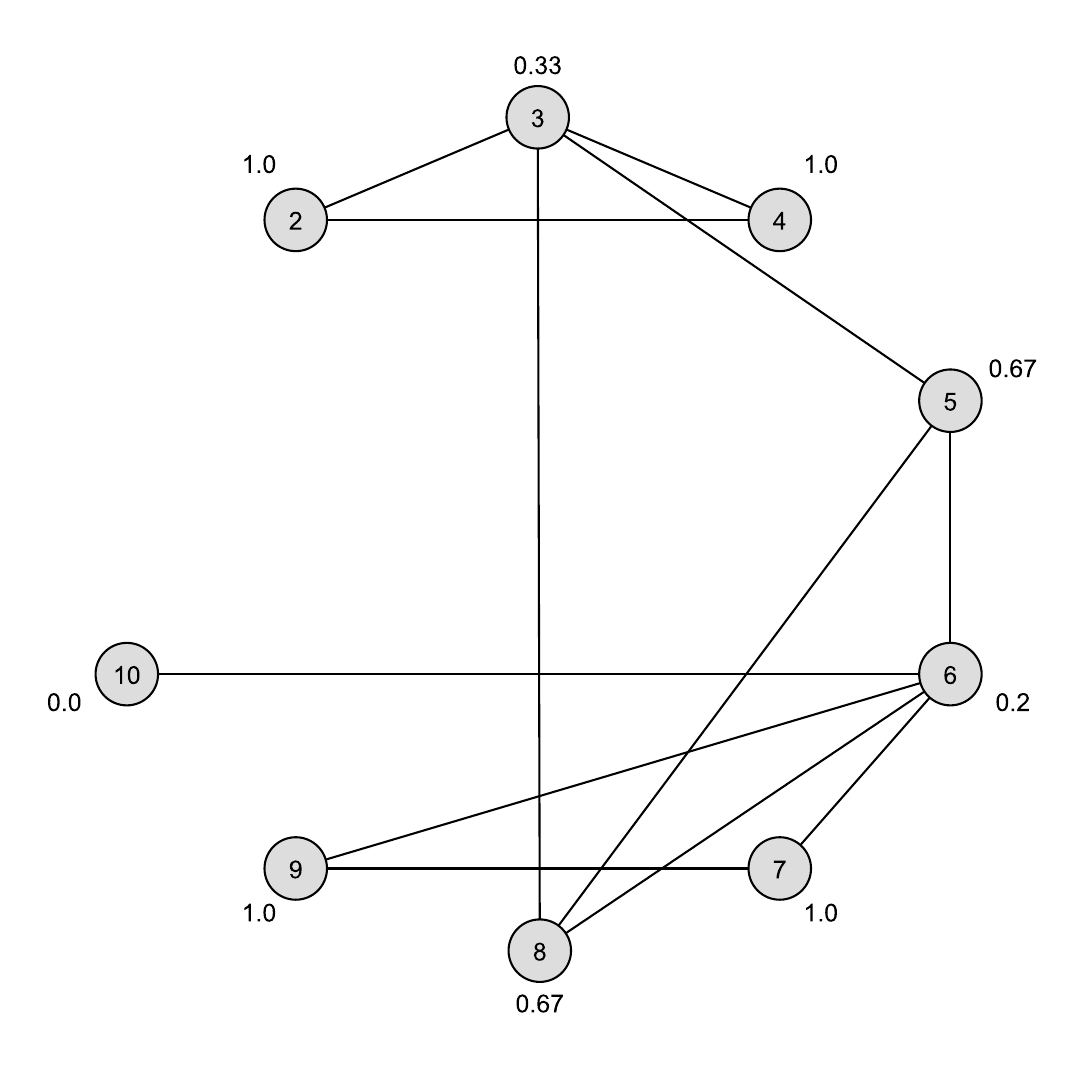}
		\caption{Undirected (\texttt{text-lcc-undirected})}
	\end{subfigure}
	\caption{Test graphs for LCC.}
	\label{fig:lcc_example}
\end{figure}

\begin{figure}[h]
	\centering
	\includegraphics[scale=\examplescale]{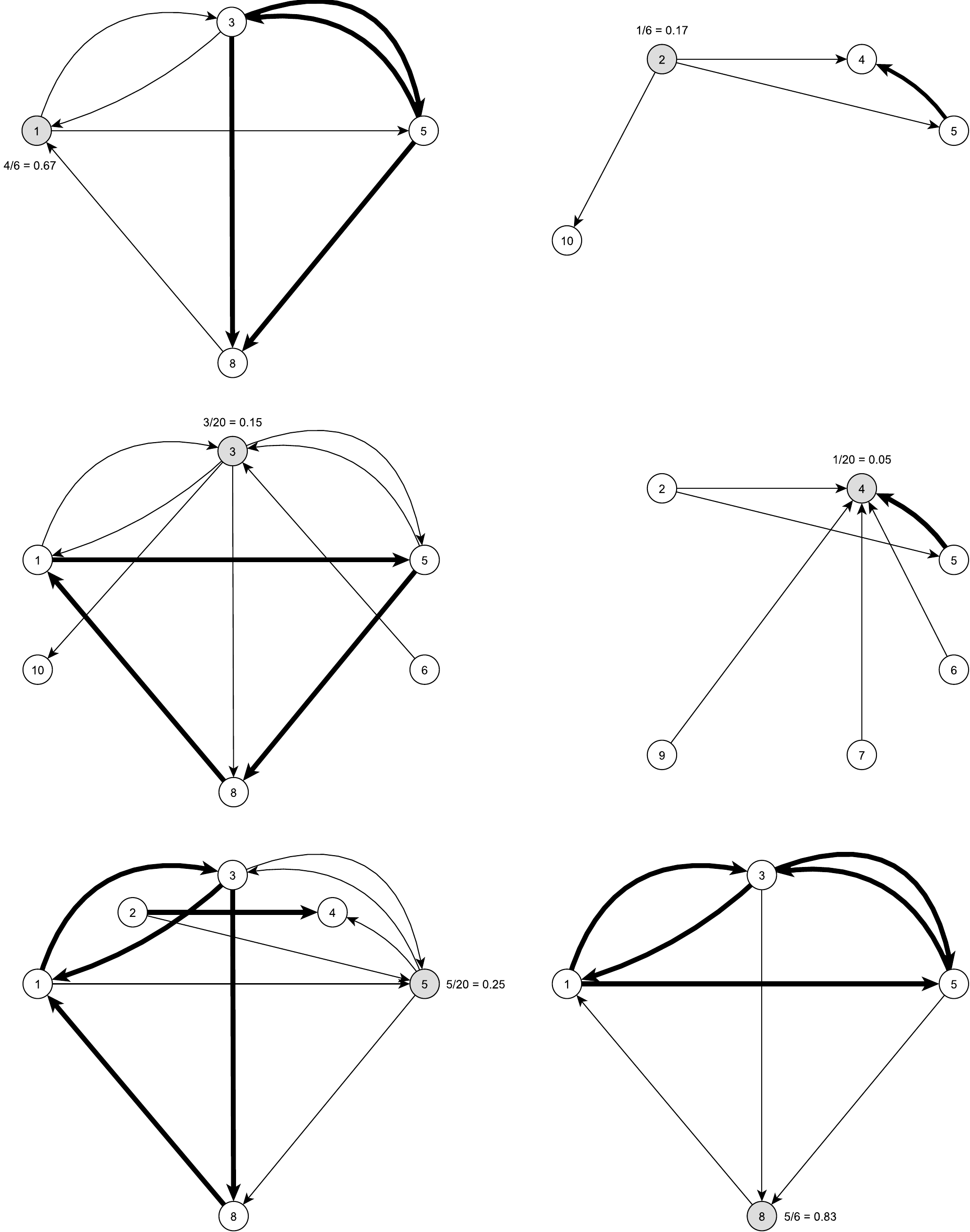}
	\caption{Detailed example of LCC values on a directed graph. Each graph represents a projected subgraph with a selected vertex (colored in gray) and its neighbors.
		Thick edges (denote the edges between the neighbors, while thin edges denote the edges from the vertex to its neighbors.
		As discussed in \autoref{sec:lcc}, the set of neighbors is determined without taking directions into account, but each neighbor is only counted once. However, directions are enforced when determining the number of edges (thick) between neighbors.
		Note that vertices 6, 7, 9, and 10 have an LCC value of 0.00 and are therefore omitted from the visualization.}
	\label{fig:lcc_dir_example_detailed}
\end{figure}

\begin{figure}[h]
	\centering
	\begin{subfigure}{0.496\textwidth}
		\centering
		\includegraphics[scale=\examplescale]{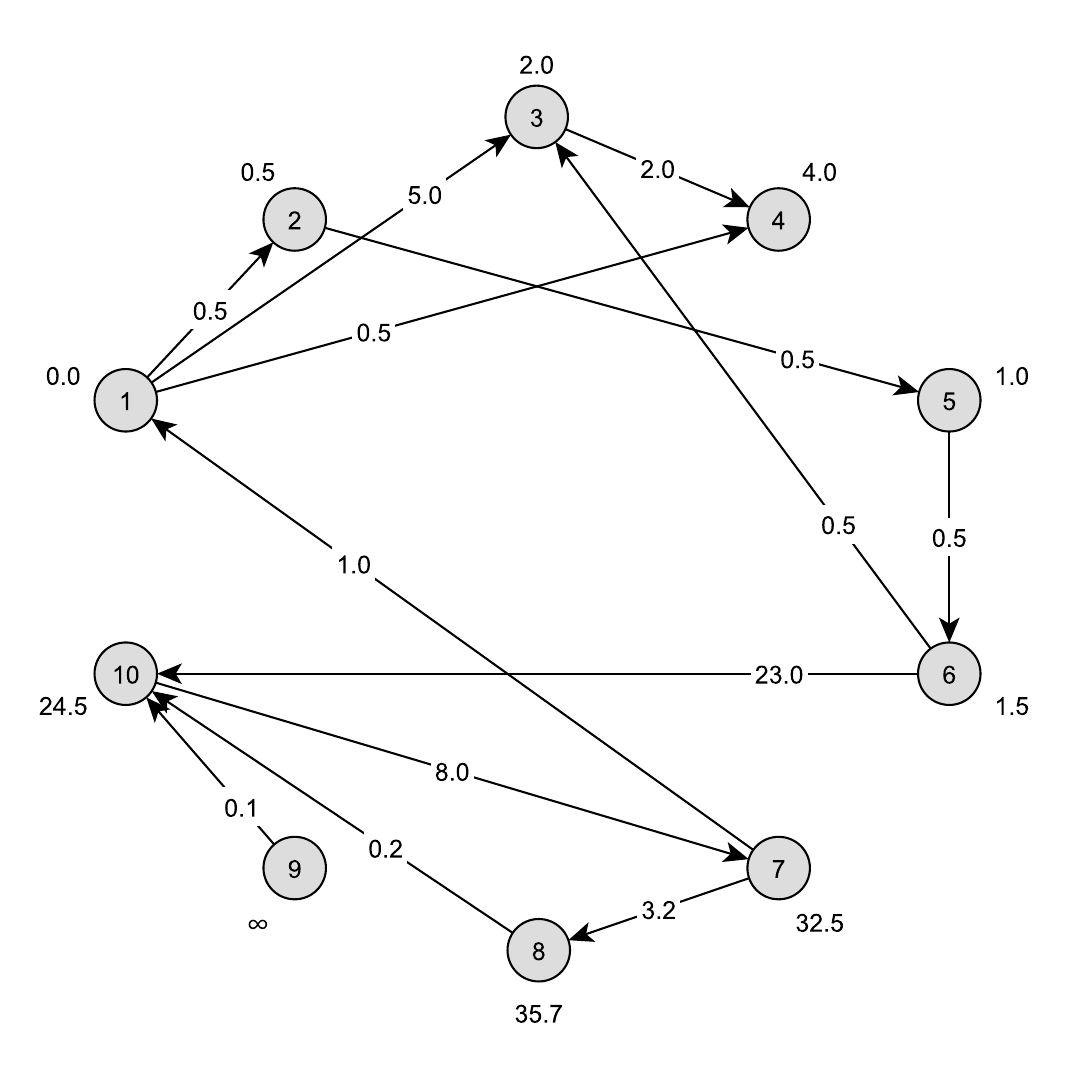}
		\caption{Directed (\texttt{text-sssp-directed})}
	\end{subfigure}
	\begin{subfigure}{0.496\textwidth}
		\centering
		\includegraphics[scale=\examplescale]{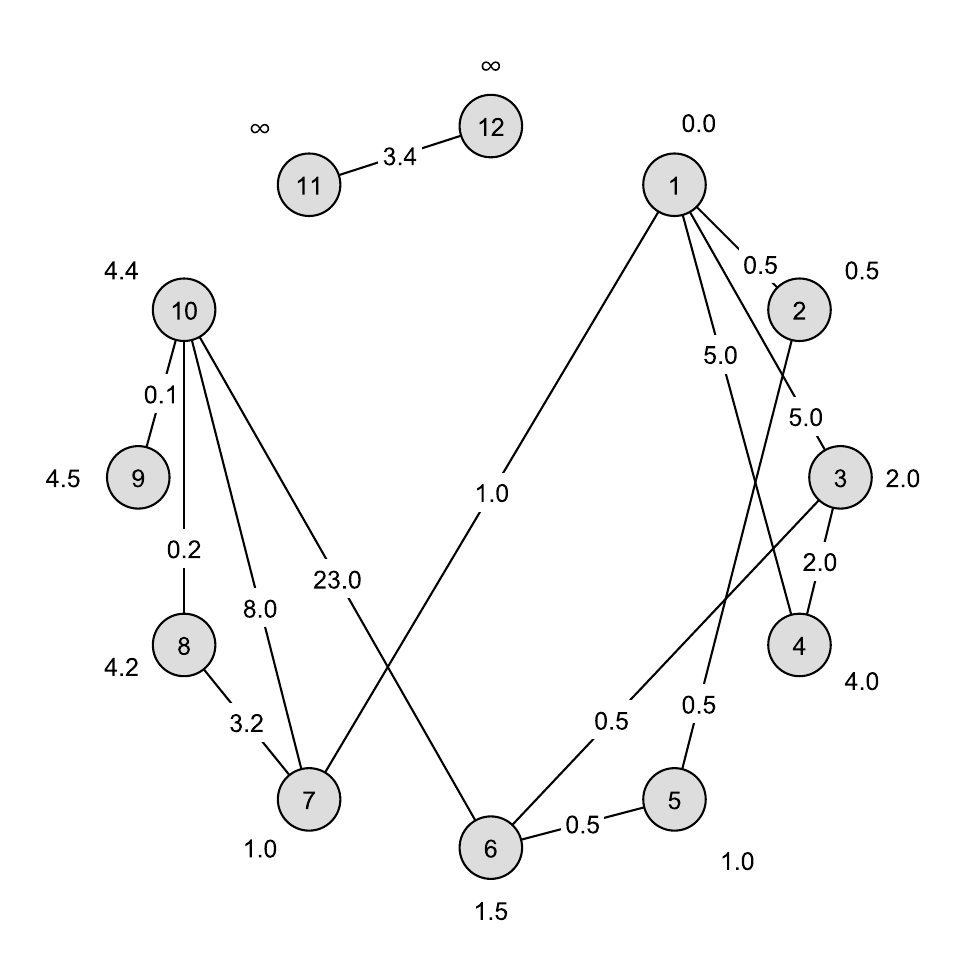}
		\caption{Undirected (\texttt{text-sssp-undirected})}
	\end{subfigure}
	\caption{Test graphs for SSSP. Source vertex: 1.}
	\label{fig:sssp_example}
\end{figure}

\begin{figure}[h]
	\centering
	\begin{subfigure}[t]{0.496\textwidth}
		\centering
		\includegraphics[scale=\examplescale]{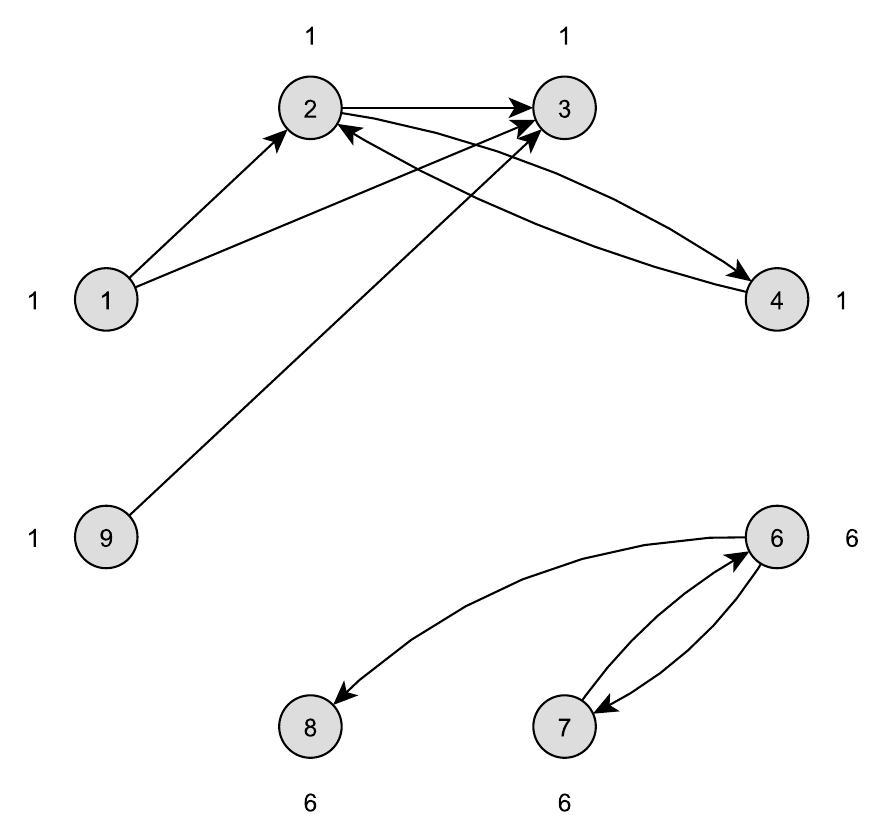}
		\caption{Directed (\texttt{text-wcc-directed}). Note that due to the semantics of the WCC algorithm, the direction of the edges is not taken into account, i.e., they are treated as undirected edges.}
	\end{subfigure}
	\begin{subfigure}[t]{0.496\textwidth}
		\centering
		\includegraphics[scale=\examplescale]{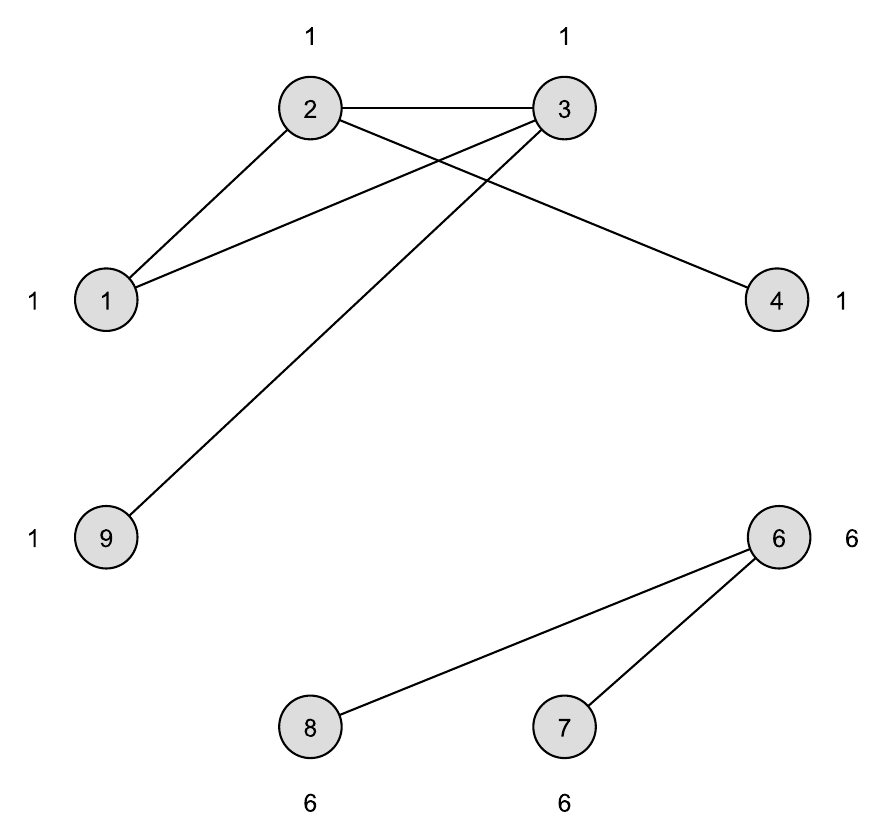}
		\caption{Undirected (\texttt{text-wcc-undirected})}
	\end{subfigure}
	\caption{Test graphs for WCC. Remark: there are 8 nodes indexed between 1 and 9 but number 5 is not assigned to any of the nodes.}
	\label{fig:wcc_example}
\end{figure}

\begin{figure}[h]
	\centering
	\begin{subfigure}{\textwidth}
		\centering
		\includegraphics[scale=\examplescale]{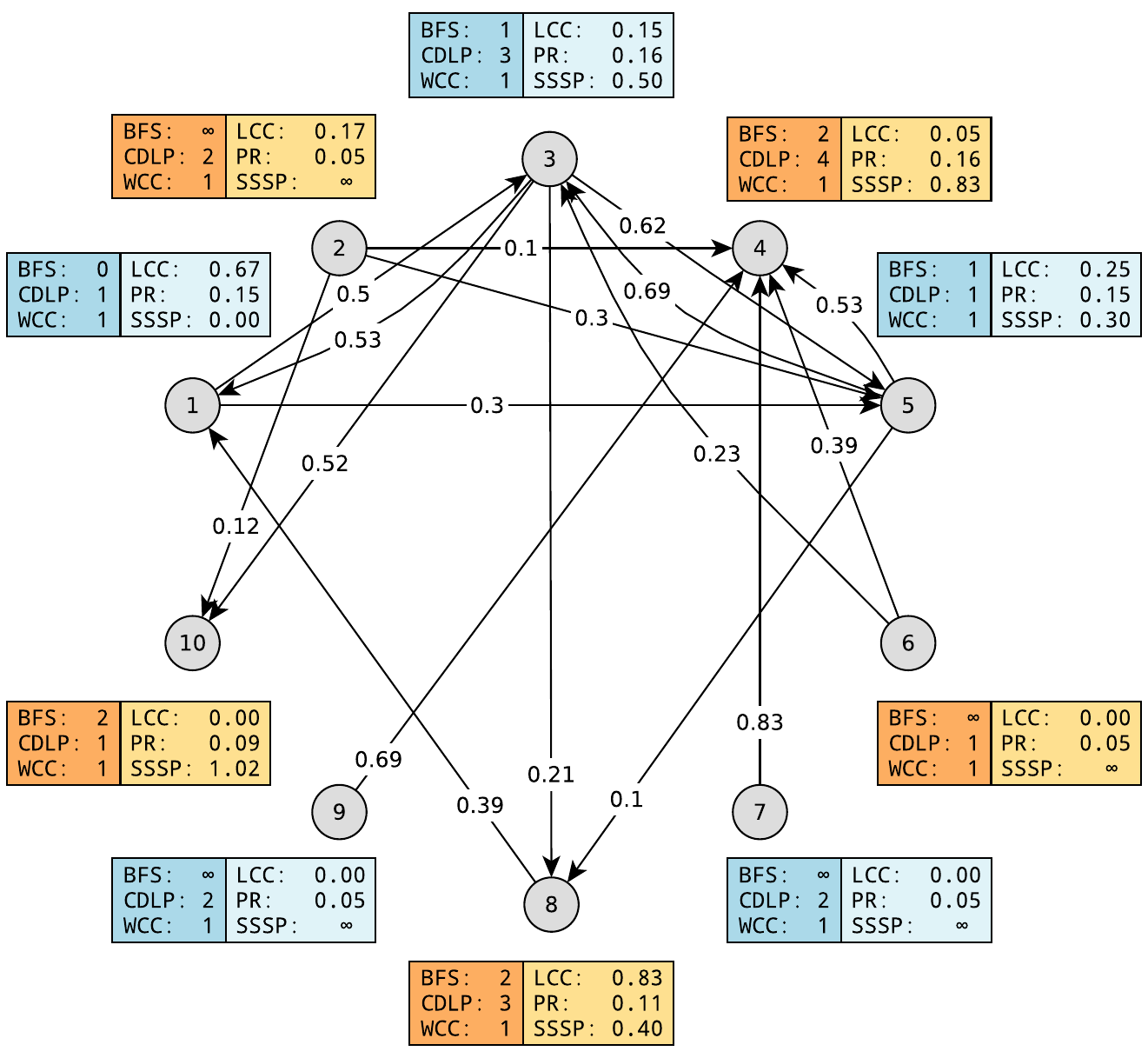}
		\caption{Directed (\texttt{example-directed}).
			Algorithm parameters --
			BFS: source vertex = 1.
			CDLP: maximum number of iterations = 2.
			PR: $d = 0.85$, maximum number of iterations = 2.
			SSSP: source vertex = 1.}
	\end{subfigure}
	\begin{subfigure}{\textwidth}
		\centering
		\includegraphics[scale=\examplescale]{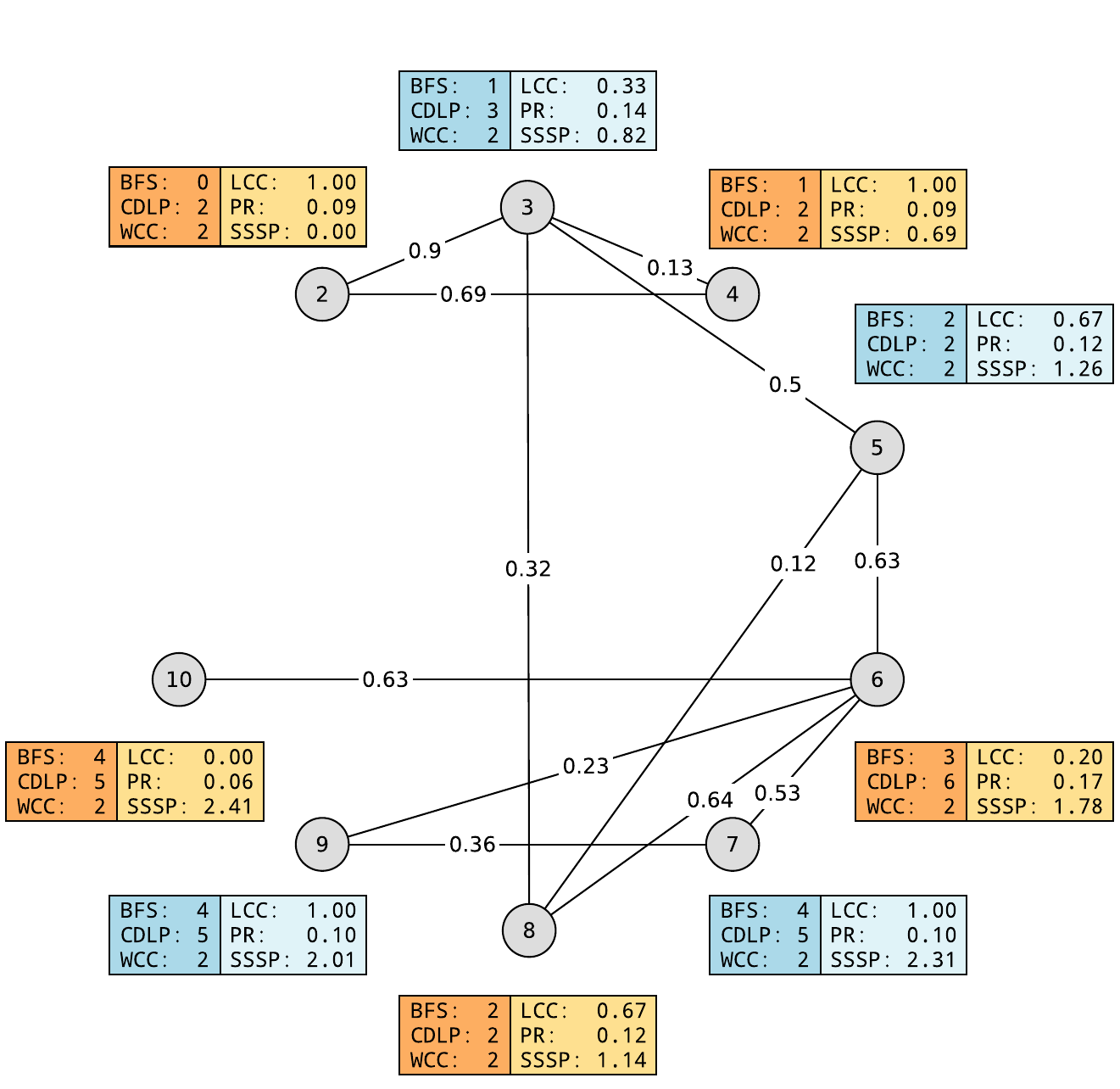}
		\caption{Undirected (\texttt{example-undirected}).
			Algorithm parameters --
			BFS: source vertex = 2.
			CDLP: maximum number of iterations = 2.
			PR: $d = 0.85$, maximum number of iterations = 2.
			SSSP: source vertex = 2.}
	\end{subfigure}
	\caption{Common examples for the six core graph algorithms.}
	\label{fig:common_example}
\end{figure}

\end{document}